\documentclass[twocolumn,floats,floatfix,aps,pra]{revtex4-2}
\usepackage{psfrag,graphicx}
\usepackage{bm,color}
\usepackage{latexsym,multirow}
\usepackage{epstopdf}
\usepackage[english]{babel}
\usepackage{amsmath,amssymb,amsfonts}
\usepackage{appendix}
\usepackage{bbold}
\usepackage[dvipsnames]{xcolor}

\usepackage{enumitem} 

\usepackage{xcolor} 

\usepackage{float}
\usepackage{dblfloatfix}

\usepackage[caption=false]{subfig}
\usepackage{comment}

\usepackage[caption=false]{subfig}

\definecolor{mygrey}{gray}{0.35}
\definecolor{myblue}{rgb}{0.2,0.2,0.8}
\definecolor{myzard}{cmyk}{0,0,0.05,0}
\definecolor{mywhite}{rgb}{1,1,1}
\definecolor{mywhite}{rgb}{1,1,1}
\definecolor{myred}{rgb}{1,0.,0.3}

\usepackage[colorlinks=true, linkcolor=blue, citecolor=blue, urlcolor=blue]{hyperref}

\def\be{ \begin{equation}}
\def\ee{ \end{equation}}
\def\bse{ \begin{subequations}}
\def\ese{ \end{subequations}}
\def\bea#1\ea{\begin{align}#1\end{align}}

\def\bi{\begin{itemize}}
\def\ei{\end{itemize}}

\def\bt{\begin{tabular}}
\def\et{\end{tabular}}

\def\ket#1{\vert #1 \rangle}

\def\3half{\tfrac32}

\def\to{\rightarrow}

\def\i{\textrm{i}}

\def\fac{\kappa}
\def\ev{\eta}

\begin{document}

\author{Julian K. Dimitrov}
\email{yulian@uni-sofia.bg}
\affiliation{Center for Quantum Technologies, Department of Physics, Sofia University, James Bourchier 5 blvd., 1164 Sofia, Bulgaria}

\author{Nikolay V. Vitanov}
\affiliation{Center for Quantum Technologies, Department of Physics, Sofia University, James Bourchier 5 blvd., 1164 Sofia, Bulgaria}

\title{Optimization of multistate stimulated Raman adiabatic passage by pulse shaping}
\date{\today }

\begin{abstract}
We propose a pulse-shaping method for improving population transfer in multistate stimulated Raman adiabatic passage (STIRAP) with nearest-neighbour couplings only. 
The method extends the quasiparallel-eigenenergy approach, previously used for two- and three-state systems, to multistate chains without introducing additional shortcut fields. 
The pump and Stokes fields are shaped so that the eigenenergies closest to the dark state remain nearly parallel to it during the central part of the interaction, where nonadiabatic transitions are most likely to occur. 
We derive analytic prescriptions for the required pulse shapes and apply them to five-, seven-, and nine-state chains formed by magnetic sublevels of degenerate manifolds with angular momenta $J_g\rightarrow J_e=J_g$ and $J_g\rightarrow J_e=J_g-1$ driven by a pair of right- and left-circularly polarized pulses. 
Numerical simulations show that the optimized pulses can reduce the population transfer error by several orders of magnitude compared to standard Gaussian pulses. 
The same shapes also improve robustness against variations of the peak Rabi frequency, the single-photon detuning, and the multiphoton detuning. 
The comparison between the two angular-momentum families shows that the $J_g\rightarrow J_g$ chains are generally more favourable, because their Clebsch-Gordan coefficients keep the relevant bright-state gap larger in the region where the dark state changes most rapidly. 
Finally, simulations with spontaneous emission show that the optimized resonant pulses remain advantageous in the lossy regime as well.
\end{abstract}

\maketitle


\section{\label{sec:intro}Introduction}

Stimulated Raman adiabatic passage (STIRAP) is one of the major methods for coherent population transfer in quantum systems \cite{Kuklinski, Vitanov-Rangelov}. 
In its simplest form, it transfers population between the two end states of a three-state chain, $|1\rangle\rightarrow |2\rangle \rightarrow |3\rangle$, by means of two delayed laser pulses \cite{Gaubatz}. 
The pump pulse couples the initial state $|1\rangle$ to the intermediate state $|2\rangle$, while the Stokes pulse couples state $|2\rangle$ to the target state $|3\rangle$. 
The pulses are applied in counterintuitive order, with the Stokes pulse preceding the pump pulse. 
When the two end states are kept on two-photon resonance, the Hamiltonian has a null-eigenvalue dark eigenstate which connects the initial and target states and has no component from the middle state. 
If the evolution is sufficiently adiabatic, the system follows this dark state and the transfer is completed with little or no population in the intermediate state at any time.

This dark-state mechanism makes STIRAP robust, but it does not remove the need for a sufficiently large pulse area. 
Very high transfer efficiency usually requires strong and/or long pulses, especially when the system contains many coupled states (as it happens in multistate STIRAP) and the adiabatic spectrum becomes dense. 

Several strategies have been developed to accelerate or improve adiabatic passage. 
Composite STIRAP sequences increase robustness by repeating the transfer with controlled phases \cite{Torosov2013}. 
Shortcut-to-adiabaticity methods add auxiliary couplings designed to cancel nonadiabatic transitions \cite{Guery-Oderin-revModPhys-STA, Unanyan}; in multistate chains, this can require several additional fields \cite{Vitanov-STA-multiSTIRAP-2017}. 
Pulse shaping provides a less intrusive alternative: one modifies only the pump and Stokes fields already present in ordinary STIRAP, without adding direct couplings between the initial and target states.

The pulse-shaping strategy used here is based on the idea of quasiparallel eigenenergies. 
Its motivation comes from the Dykhne-Davis-Pechukas approach to nonadiabatic transitions \cite{Dykhne, Davis-Pechukas}. 
In that formulation, the transition probability between adiabatic eigenstates is controlled by the analytic structure of the eigenenergy splitting in the complex-time plane. 
A slowly varying and nearly constant eigenenergy gap reduces the probability of leaving the desired adiabatic state. 
For finite pulses, exact parallelism of eigenenergies cannot be maintained at all times, because the fields must vanish at the beginning and at the end of the interaction as their energy is finite.

In multistate systems, it is impossible to make all eigenvalues parallel, even for a limited time.
Fortunately, this is not necessary because nonadiabatic leakage from the dark state is most likely to occur to its closest neighbours.
Therefore, we demand that the eigenenergy closest to the dark-state energy (i.e. zero) should be kept nearly parallel in the central part of the pulse, where the mixing angle changes most rapidly and nonadiabatic leakage is most likely.

This approach has already proved useful in two- and three-state systems. 
A rigorous construction was given for two-state dynamics in Ref.~\cite{Stefan-2lvl-parallel}. 
In STIRAP, Vasilev \textit{et al.} showed that the Dykhne-Davis-Pechukas argument leads to the same flat-gap condition both on resonance and in the large-detuning limit \cite{Vasilev2009}. 
Dridi \textit{et al.} considered fully parallel eigenenergies by allowing a nonzero two-photon detuning \cite{DridiParallel}, while Liu \textit{et al.} optimized the fields by constraining the integrated population of the lossy intermediate state \cite{Guerin2023}. 
These results show that the shape of the adiabatic spectrum, not only the pulse area, can be used as a control resource.
However, these two latter works placed sizable transient population in the middle state, which could be detrimental if this is a lossy state. 
On the contrary, the quasiparallel pulse shaping approach used by us keeps the lossy middle state virtually unpopulated.

The present work extends this idea to multistate STIRAP chains with an odd number of states. 
Adiabatic transfer in multistate chains was investigated soon after the introduction of three-state STIRAP \cite{Marte1991,Shore1991,Smith1992}; a comprehensive review is given in Ref.~\cite{Vitanov-Rangelov}. 
On multiphoton resonance, a chain with $N=2n+1$ states possesses a zero-eigenvalue eigenstate containing only the lower-chain states \cite{Marte1991,Shore1991,Smith1992,Vitanov1998}. 
With counterintuitively ordered couplings, this state connects the two ends of the chain while avoiding the even-numbered, typically lossy, intermediate states. 
Chains with an even number of states do not generically possess the same resonant end-to-end dark-state connection and require additional constraints on the couplings and the detunings \cite{Shore1992Comparative,Vitanov1998}. 
We therefore restrict attention to chains with an odd number of states, where the dark-state mechanism is generic and native.

Multilevel dark-state transport is relevant in a wide range of physical settings. 
In atoms, it has been demonstrated between magnetic sublevels and used to prepare coherent superpositions of several Zeeman states \cite{Pillet1993,Martin1996,Vewinger2003,Heinz2006}. 
In atom optics, chains of internal and momentum states have been used for coherent momentum transfer, atomic mirrors, beam splitters, and interferometers \cite{Marte1991,Goldner1994a,Goldner1994b,Weitz1994a,Weitz1994b}. 
Collective internal and vibrational states of trapped-ion chains can likewise form multistate STIRAP linkages for the preparation of Dicke states \cite{Linington2008a,Noguchi2012}. 
In molecular physics, STIRAP has become an essential tool for transferring weakly bound ultracold molecules to deeply bound rovibrational levels \cite{Lang2008,Danzl2008,Ni2008,Danzl2010}. 

Here we focus on chainwise systems formed by the magnetic sublevels of two degenerate manifolds with angular momenta $J_g$ and $J_e$ driven by a pair of right- and left-circularly polarized pulses that drive only transitions with $\Delta m=\pm 1$. 
The two cases considered are $J_e=J_g$ and $J_e=J_g-1$, with three-, five-, seven-, and nine-state examples. 
The two pulse envelopes are shared by all couplings in the chain, while the individual couplings differ by their Clebsch-Gordan coefficients. 
This makes the comparison between the two angular-momentum families especially transparent: the different transfer efficiencies can be traced directly to the different eigenenergy gaps produced by these coefficients.

The main result is an analytic pulse-shaping prescription for the pump and Stokes fields. 
The dark-state evolution is fixed by a monotonic mixing function, while a common parallelizing function reshapes the smallest eigenenergy to remain nearly constant during the main part of the evolution. 
We derive the corresponding functions for three- to nine-state chains and test them numerically. 
The optimized pulses reduce the transfer error by several orders of magnitude compared with standard Gaussian pulses and also broaden the high-efficiency region with respect to the peak Rabi frequency, the single-photon detuning, and the multiphoton detuning.
Again, optimized pulses which generate quasiparallel eigenenergies, outperform standard Gaussian pulses.

The paper is organized as follows. 
Section~\ref{sec:theory} reviews three-state STIRAP and introduces the quasiparallel pulse-shaping approach. 
Section~\ref{subsec:MultilevelSTIRAP} extends the formalism to multistate chains formed by an odd number of degenerate magnetic sublevels and discusses the numerical results for five-, seven-, and nine-state systems. 
Section~\ref{sec:dissipations} studies spontaneous emission in three- and five-state examples. 
Section~\ref{sec:conclusion} summarizes the results. 
The explicit algebra leading to the parallelizing functions is presented in Appendix~\ref{appendix}. Sensitivity to field noise is studied in \ref{app:noisy_pulses} and \ref{app:dephasing} covers the systems’ susceptibility to dephasing.


\section{\label{sec:theory} Three-state STIRAP}

\subsection{\label{subsec:3lvlSTIRAP} Standard STIRAP}

We describe the three-state STIRAP dynamics in terms of the dimensionless time $\tau=t/T$ with an arbitrary reference timescale $T$.
Equivalently, the physical Rabi frequencies and detunings are multiplied by the time scale $T$; in what follows we keep the same symbols $\Omega_p(\tau)$, $\Omega_s(\tau)$ and $\Delta(\tau)$ for these dimensionless quantities. 
Throughout the paper, $\hbar=1$. The probability amplitudes $\textbf{c}(\tau)=[c_1(\tau),c_2(\tau),c_3(\tau)]^T$ then obey the Schr\"odinger equation,
\begin{equation}
 i\frac{\partial}{\partial \tau}\textbf{c}(\tau)=\textbf{H}(\tau)\textbf{c}(\tau),
\end{equation}
where, under the rotating-wave approximation and in the absence of dissipation, the Hamiltonian is

\begin{equation}
\textbf{H}(\tau)=\frac{1}{2}
\left[\begin{matrix}
 -\delta(\tau) & \Omega_p(\tau) & 0 \\
 \Omega_p(\tau) & 2\Delta(\tau) & \Omega_s(\tau) \\
 0 & \Omega_s(\tau) & \delta(\tau)
\end{matrix}\right].
\label{eq:ham_3lvl}
\end{equation}

\begin{figure}[tbph]
\begin{tabular}{r}
\centering
 \includegraphics[width=0.65\linewidth]{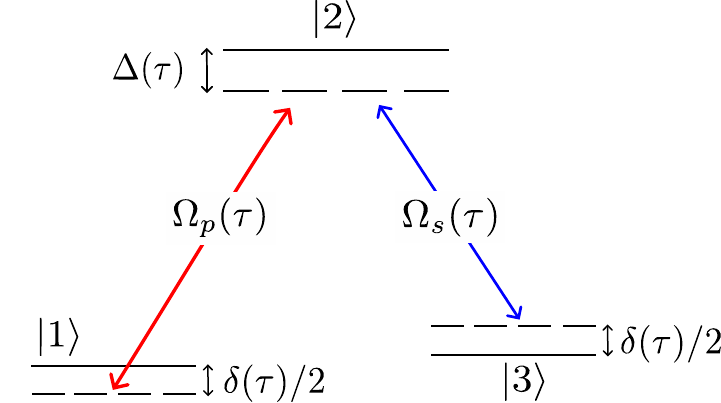}
\end{tabular}
\caption{Three-state $\Lambda$ configuration for STIRAP on two-photon resonance. 
The pump field $\Omega_p(\tau)$ couples the initial state $|1\rangle$ to the intermediate state $|2\rangle$, while the Stokes field $\Omega_s(\tau)$ couples state $|2\rangle$ to the target state $|3\rangle$. 
The intermediate state has the single-photon detuning $\Delta(\tau)$.}
\label{fig:3lvlSTIRAP}
\end{figure}

Figure \ref{fig:3lvlSTIRAP} shows the corresponding $\Lambda$ configuration. 
The pump field $\Omega_p(\tau)$ couples the $|1\rangle-|2\rangle$ transition, whereas the Stokes field $\Omega_s(\tau)$ couples the $|2\rangle-|3\rangle$ transition; both Rabi frequencies can be taken real without loss of generality. 
The single-photon detuning $\Delta(\tau)$ is the difference between the particular laser carrier frequency and the corresponding Bohr transition frequency.

The two-photon detuning $\delta(\tau)$ is the difference between the net energy delivered by driving laser fields and the actual energy gap between the initial state $|1\rangle$ and the final target state $|3\rangle$. 
It can arise from calibration error in the driving pulses or from external magnetic or electric fields altering the energy gaps.
A carrier laser frequency mismatch can shift each field frequency independently, if derived from different sources, or simultaneously, if derived from the same source.
In the former case both $\Delta$ and $\delta$ would shift; in the latter, only $\Delta$ would change.
Detunings can also arise due to external magnetic or electric fields due to the linear Zeeman and quadratic Stark effects.
The linear Zeeman shift is proportional to the magnetic quantum number $m$; hence if states $\ket{1}$ and $\ket{3}$ are the magnetic sublevels with $m=-1$ and $+1$, respectively, as it is often the case, then the two-photon detuning $\delta$ would be directly proportional to the magnetic field. 
In that case the upper middle state $\ket{2}$ would have $m=0$ and would not experience a magnetic shift. 
The Stark shift is proportional to $m^2$ and hence it would not cause two-photon shift in such an exemplary system. 
In the analytic considerations below we will assume that $\delta=0$, unless stated otherwise. In the simulations in the figures we will include the two-photon detuning explicitly.

The zero off-diagonal corners indicate that no direct field is applied on the $|1\rangle-|3\rangle$ transition, which in atoms and ions may be electric-dipole forbidden.
Indeed, it is a major advantage of the present pulse-shaping approach that it avoids the difficulty in driving this forbidden transition, which is demanded by the shortcuts-to-adiabaticity approach.
We keep a nonzero single-photon detuning in the model in order to examine its effect on the transfer efficiency.

The central feature of STIRAP is the time-dependent dark state, i.e. the zero-eigenvalue adiabatic state of the Hamiltonian,
\begin{equation}
 |d\rangle=\cos\theta(\tau)|1\rangle-\sin\theta(\tau)|3\rangle .
\end{equation}
Its probability amplitude is correspondingly
\begin{equation}
 d(\tau)=\cos\theta(\tau)c_1(\tau)-\sin\theta(\tau)c_3(\tau).
\end{equation}
The mixing angle is defined by
\be
\tan\theta(\tau)=\Omega_p(\tau)/\Omega_s(\tau).
\ee
For counterintuitively ordered pulses, with the Stokes pulse preceding the pump pulse, $\theta(\tau)$ changes from $0$ to $\pi/2$. Thus the dark state coincides with the initial state $|1\rangle$ at the beginning and with the target state $|3\rangle$ at the end, up to a minus sign.

The transfer is complete in the adiabatic limit, where transitions from the dark state to the two bright adiabatic states are negligible. We denote differentiation with respect to $\tau$ by a dot. The usual adiabatic conditions can then be written as
\be
\left|\dot\theta\frac{\sin^2\phi}{\cos\phi}\right|\ll \tfrac12\Omega,\quad
\left|\dot\theta\frac{\cos^2\phi}{\sin\phi}\right|\ll \tfrac12\Omega,
\ee
where
\be\label{phi}
\tan 2\phi(\tau)=\frac{\Omega(\tau)}{\Delta(\tau)},\quad
\Omega(\tau)=\sqrt{\Omega_p(\tau)^2+\Omega_s(\tau)^2}.
\ee
On resonance, $\Delta=0$ and $\phi=\pi/4$, so the adiabatic condition reduces to
\begin{equation}
 |\sqrt{2}\,\dot\theta|\ll \Omega .
\end{equation}
After integration over the interaction time, this gives the familiar requirement of a large pulse area, $A\gg1$, up to pulse-shape factors.

\subsection{\label{subsec:Opt3lvlSTIRAP}Optimization of three-state STIRAP}


\begin{figure}[tbph]
 \centering
 \setlength{\tabcolsep}{6pt}
 \renewcommand{\arraystretch}{1.0}
 \includegraphics[width=0.9\linewidth]{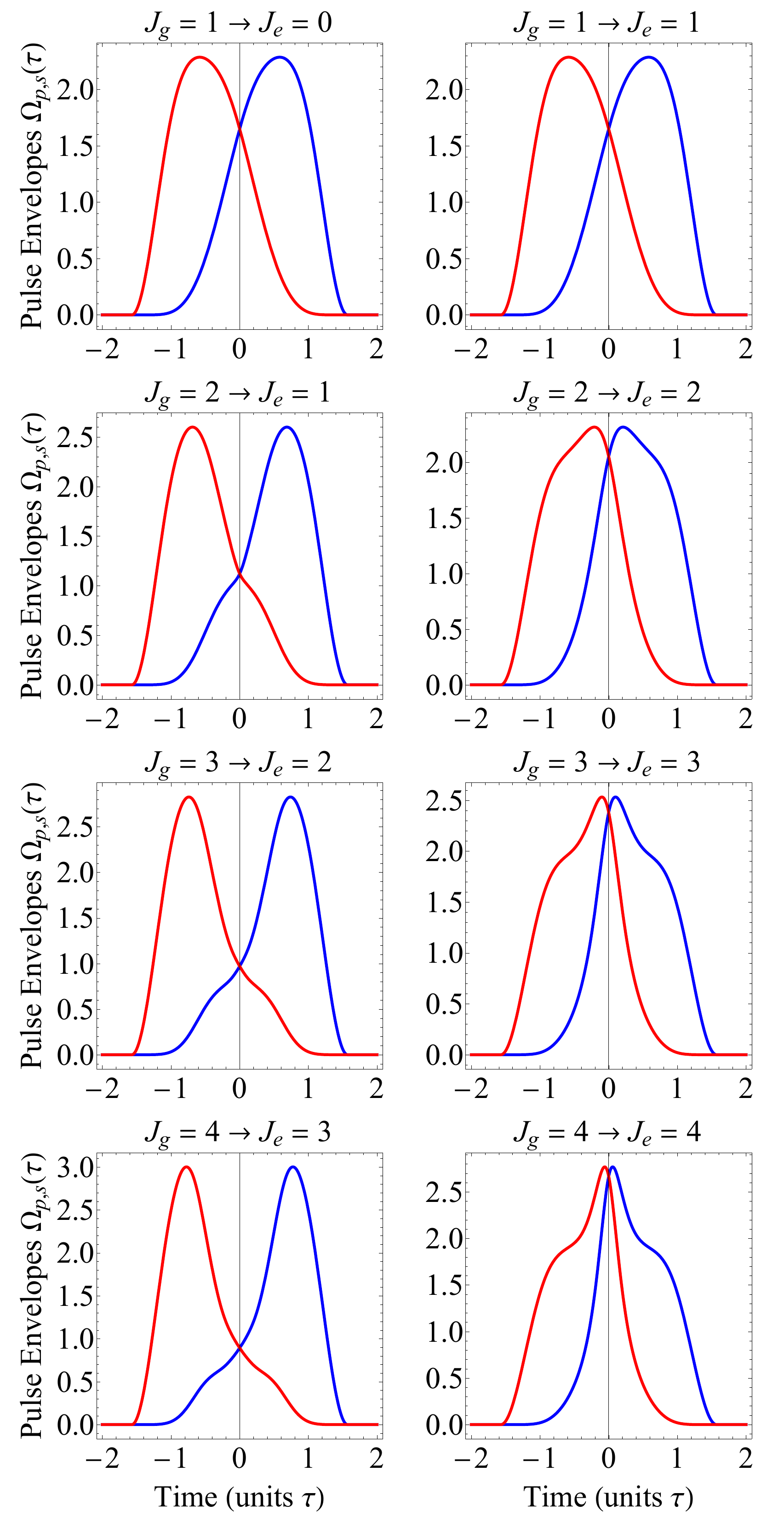}
 \caption{Normalized pump and Stokes pulse shapes for the systems considered in this paper. Each individual pulse has area $\pi$ for $\Omega_0=1$. The three-state fields are defined by Eqs.~\eqref{eq:DDPpulses} and \eqref{fg}, while the multistate fields follow Eqs.~\eqref{eq:multiPulses}, with linkage-dependent parallelizing functions derived beginning in Sec.~\ref{sec:OptMultiSTIRAP} and in Appendix~\ref{appendix}.}
 \label{fig:shapes}
\end{figure}

Optimization of STIRAP by means of quasiparallel eigenvalues for zero and very large constant single-photon detuning was proposed by Vasilev \textit{et al.} \cite{Vasilev2009}. 
The eigenvalues of the Hamiltonian on two-photon resonance are
\begin{subequations}
\begin{align}
 \ev_{-}(\tau)&=\frac{1}{2}\left[\Delta-\sqrt{\Omega_p(\tau)^2+\Omega_s(\tau)^2+\Delta^2}\right] \notag\\
 &=-\tfrac12\Omega(\tau)\tan\phi(\tau),\\
 \ev_0(\tau)&=0,\\
 \ev_{+}(\tau)&=\frac{1}{2}\left[\Delta+\sqrt{\Omega_p(\tau)^2+\Omega_s(\tau)^2+\Delta^2}\right] \notag\\
 &=\tfrac12\Omega(\tau)\cot\phi(\tau).
\end{align}
\label{eq:eigenen3lvl}
\end{subequations} 
On resonance ($\Delta=0$), we have
\be
\ev_{-}(\tau)=-\tfrac12\Omega(\tau),\quad
 \ev_0(\tau)=0,\quad
 \ev_{+}(\tau)=\tfrac12\Omega(\tau).
\label{eq:eigenen3lvl}
\ee

The quasiparallel-eigenvalue idea is to keep the relevant energy gaps as constant as possible during the time interval in which the mixing angle changes. 
Exact parallelism is incompatible with finite pulses that vanish at the beginning and at the end of the interaction, so the condition is imposed only in the central transfer region.

A convenient parametrization of the pump and Stokes pulses is
\bse
\label{eq:DDPpulses}
\begin{align}
 \Omega_p(\tau)&=\fac\Omega_0 f(\tau)\sin\theta(\tau),\\
 \Omega_s(\tau)&=\fac\Omega_0 f(\tau)\cos\theta(\tau).
\end{align}
\ese
Here $f(\tau)$ is a pulse-shaped mask which makes the fields finite, while $\theta(\tau)$ is a monotonic dimensionless mixing function satisfying $\theta(-\infty)=0$ and $\theta(+\infty)=\pi/2$. 
The factor $\fac$ is introduced in such a manner that the pulse area of each pulse is equal to $\pi$ when $\Omega_0=1$, i.e. the pulse area of each pulse in Eqs.~\eqref{eq:DDPpulses} is equal to $\pi\Omega_0$.

With this convention, $\Omega(\tau)=\fac\Omega_0 f(\tau)$.
Therefore the nonadiabatic coupling is controlled by $\dot\theta(\tau)$. 
The pulse design should consequently make $\dot \theta(\tau)$ appreciable only in the region where the eigenvalue splitting is large and nearly constant.

In the parametrization \eqref{eq:DDPpulses}
the two fields are related by time reversal and therefore have equal peak amplitudes. 
(The parameter $\Omega_0$ sets their common amplitude and pulse-area scale, but is not generally equal to the peak Rabi frequency.)
This is not mandatory, but it is the most favorable choice for STIRAP because it helps to satisfy the adiabatic condition. 
The case of unequal peak Rabi frequencies has been analyzed in Ref.~\cite{Boradjiev2013}, where it was shown that unbalanced fields deteriorate the transfer efficiency.


\begin{figure}[tbph]
 \centering
 \setlength{\tabcolsep}{6pt}
 \renewcommand{\arraystretch}{1.0}
 \includegraphics[width=0.9\linewidth]{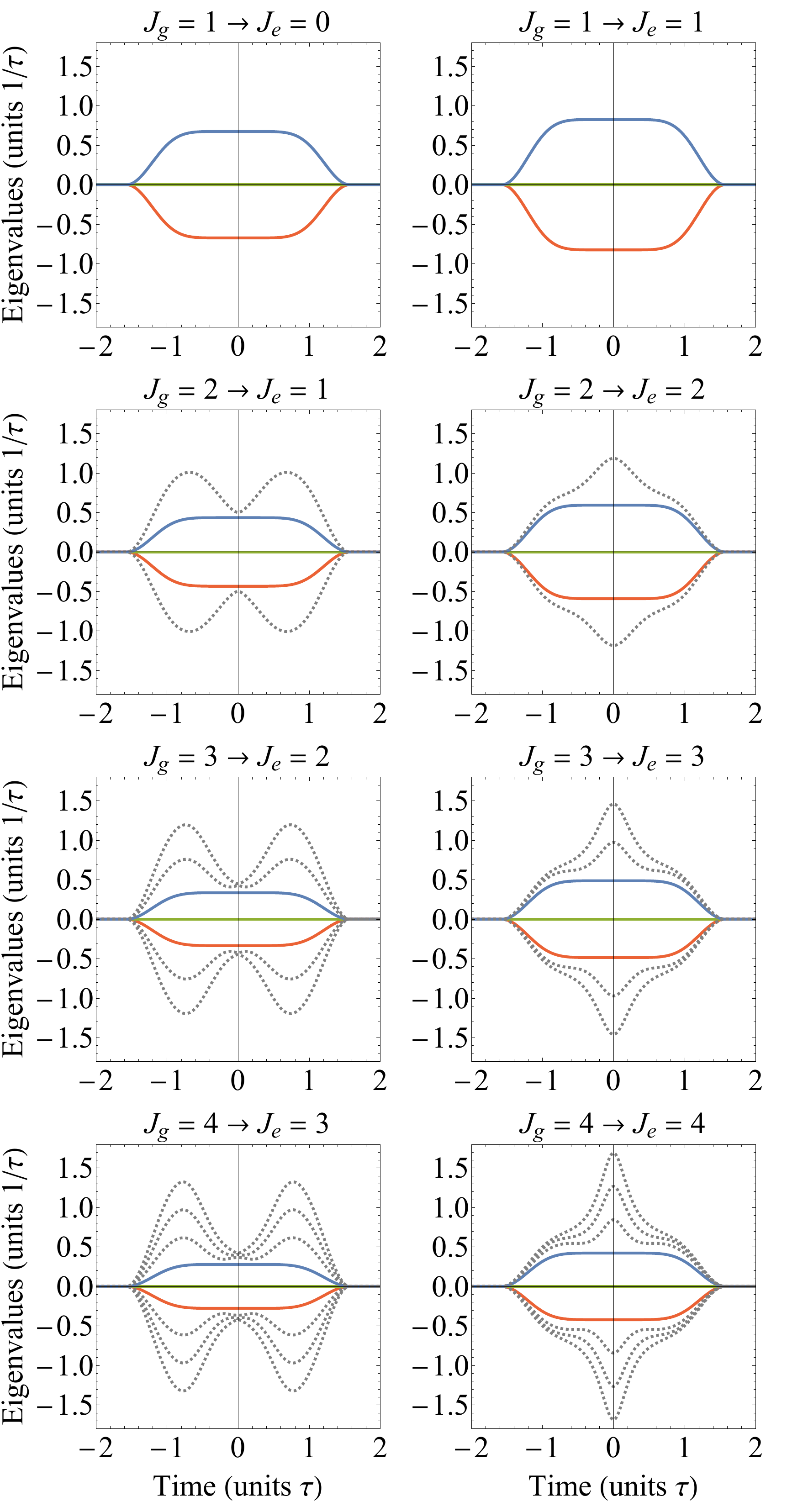}
\caption{Resonant adiabatic eigenenergies corresponding to the pulse shapes in Fig.~\ref{fig:shapes}. The systems differ only through the Clebsch-Gordan coefficients listed in Table~\ref{table:CG}, and each individual pulse has area $\pi$ for $\Omega_0=1$. The $J_e=J_g$ linkages retain a larger central dark-bright gap than the corresponding $J_e=J_g-1$ linkages, consistent with Table~\ref{tab:eigenvalue_splitting} and the transfer results in Fig.~\ref{fig:1d_rabi_res}.}
 \label{fig:eigenenergies}
\end{figure}


\begin{figure}[tbph]
 \centering
 \setlength{\tabcolsep}{6pt}
 \renewcommand{\arraystretch}{1.0}
 \includegraphics[width=1\linewidth]{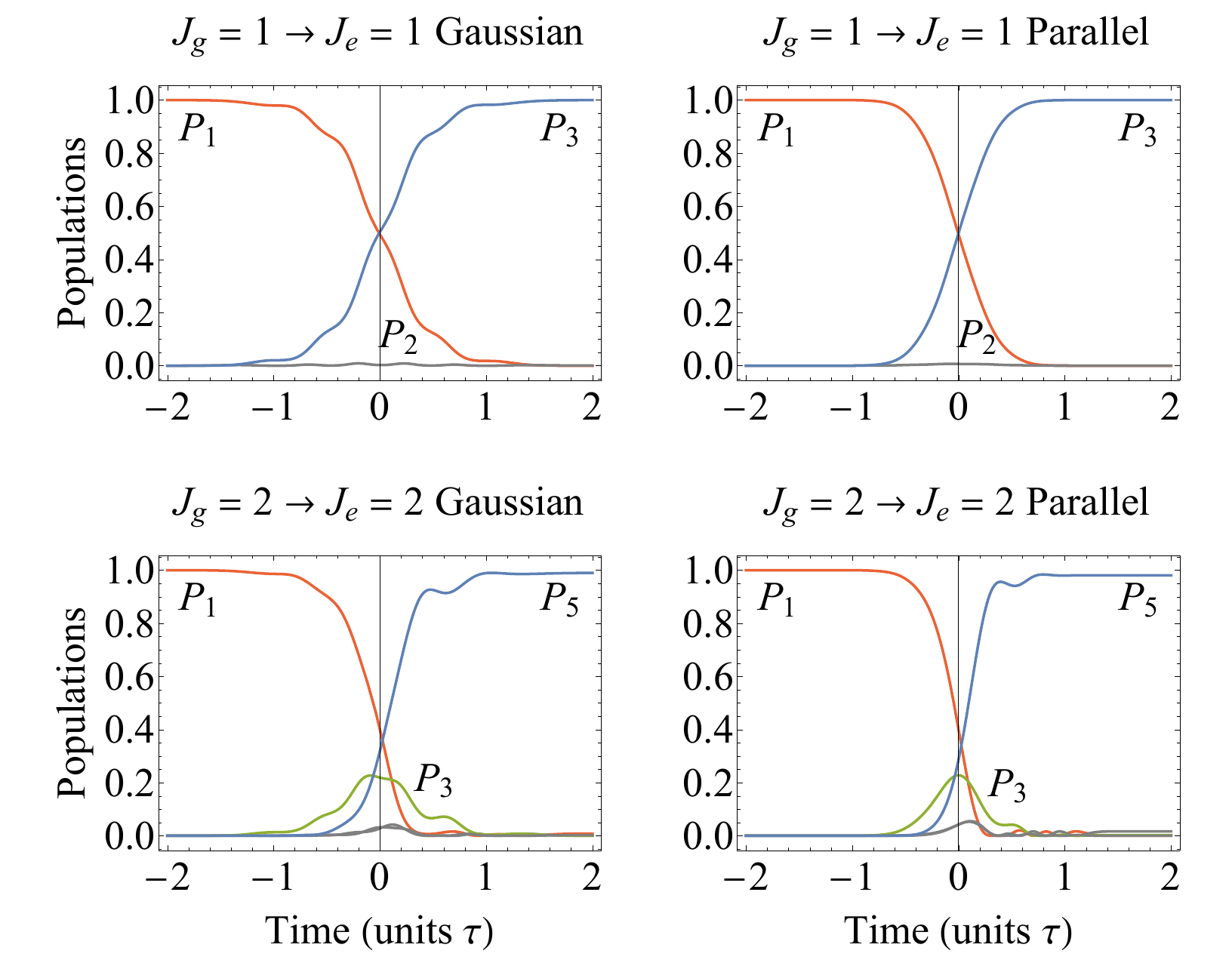}
\caption{Time evolution of the states with Gaussian pulses \eqref{eq:Gaussians} and parallel-STIRAP fields in Fig. \ref{fig:shapes}. The systems are on single-photon and multiphoton resonance with peak Rabi frequency $\Omega_0=20$. The parallel-STIRAP leaves less transient population on the excited states (gray) due to the lowered probability of non-adiabatic transition. Because of the shorter pulses compared to the Gaussians the optimized systems lead to slightly faster population transfer.}
 \label{fig:populations}
\end{figure}

In the simulations below we use
\bse\label{fg}
\label{eq:1-sin2}
\begin{align}
 f(\tau)&=1-\sin^8\tau,\\
 \theta(\tau)&=\frac{\pi}{4}+\frac{\tau}{2} + \frac{\sin(2\tau)}{3}+\frac{\sin(4\tau)}{24},
\end{align}
\ese
inside the interval $\tau\in[-\pi/2,\pi/2]$, with $f(\tau)=0$ outside this interval, $\theta(\tau)=0$ for $\tau<-\pi/2$, and $\theta(\tau)=\pi/2$ for $\tau>\pi/2$. 
The function $\theta(\tau)$ is the normalized integral of $\cos^4\tau$, so $\dot \theta(\tau)$ vanishes smoothly at the pulse edges and is concentrated near $\tau=0$. 
Thus the nonadiabatic coupling is localized in the same region in which the mask $f(\tau)$ is flat and the eigenvalue gaps are largest. 
The resulting pulse shapes are shown in Fig.~\ref{fig:shapes}.
With this choice we have for the factor $\fac \approx 2.334564$.

We deliberately choose these pulse shapes because they are both smooth and have well defined finite duration in order to facilitate implementation. 
Their performance is similar to the pulses in Ref.~\cite{Vasilev2009}, which have assumed an infinite duration.

As a basis for comparison with standard STIRAP we use Gaussian pulses,
\begin{subequations}
\label{eq:Gaussians}
\begin{align}
 \Omega_p(\tau)&=\sqrt{\pi}\, \Omega_0\exp\left[-(\tau-0.5)^2\right],\\
 \Omega_s(\tau)&=\sqrt{\pi}\, \Omega_0\exp\left[-(\tau+0.5)^2\right].
\end{align}
\end{subequations}
where the delay is chosen as a conventional near-optimal value for ordinary three-state STIRAP; it is not reoptimized separately for every multistate chain. 
These Gaussian pulses are used only as a reference case against which the quasiparallel scenario, or hereafter referred to as parallel-STIRAP, is compared.

In Fig. \ref{fig:populations} we contrast the time evolution of states for three- and five-state STIRAP for both pairs of fields. The optimized pulses lead to more stable dynamics as well as more rapid population transfer resulting from the specific choice of shapes in the current study presented in Fig. \ref{fig:shapes}.


\begin{figure}[tbph]
 \centering
 \setlength{\tabcolsep}{6pt}
 \renewcommand{\arraystretch}{1.0}
 \includegraphics[width=0.9\linewidth]{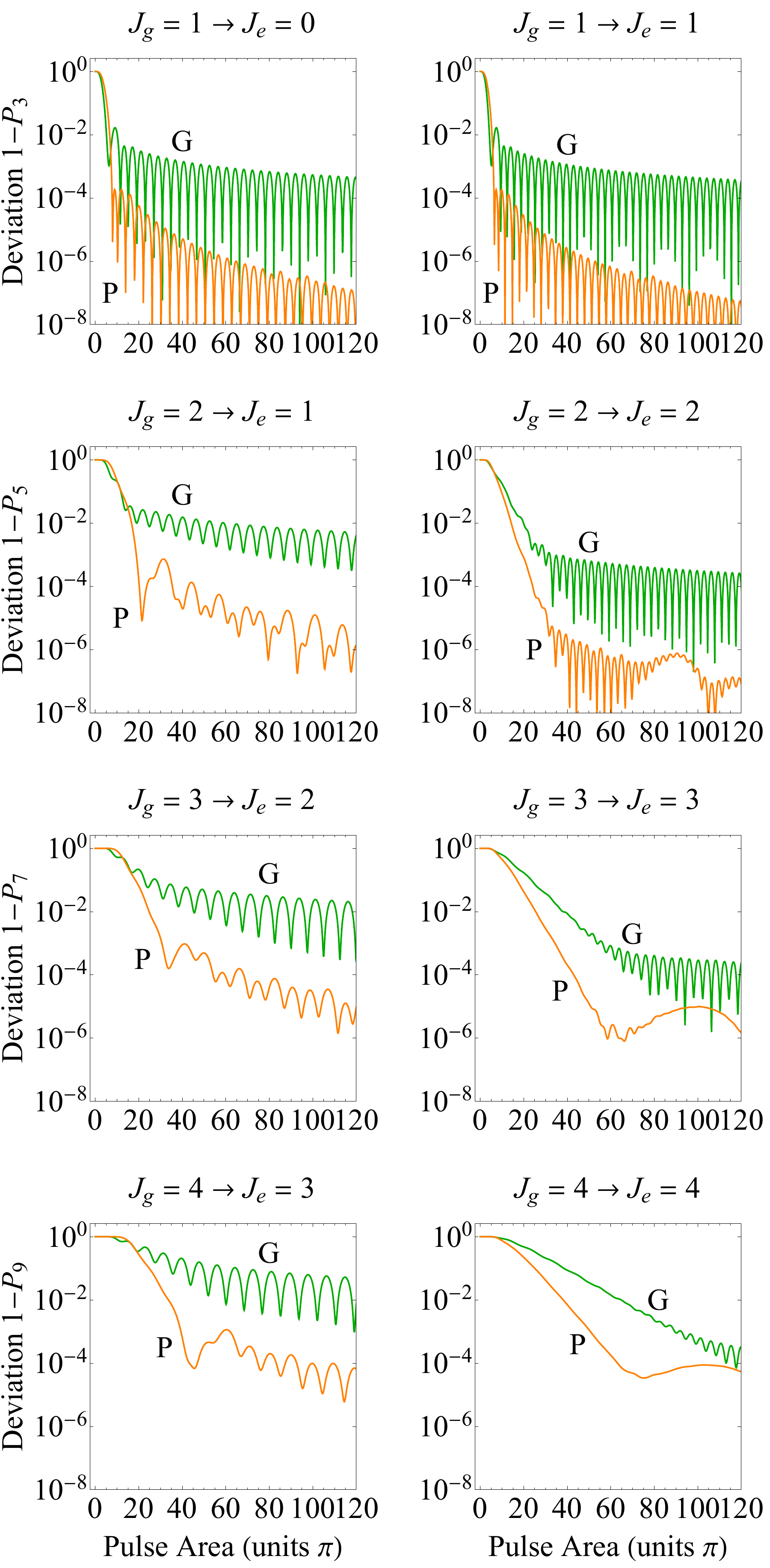}
\caption{Population-transfer error $1-P_N$ as a function of the temporal pulse area for the resonant Gaussian and parallel-STIRAP pulses. The optimized pulses produce lower errors for all chain lengths, while the $J_e=J_g$ linkages generally outperform the corresponding $J_e=J_g-1$ linkages because of their larger dark-bright gaps. The labels G and P denote Gaussian and parallel-STIRAP pulses, respectively.}
 \label{fig:1d_rabi_res}
\end{figure}


\begin{figure}[tbph]
 \centering
 \setlength{\tabcolsep}{6pt}
 \renewcommand{\arraystretch}{1.0}
 \includegraphics[width=0.9\linewidth]{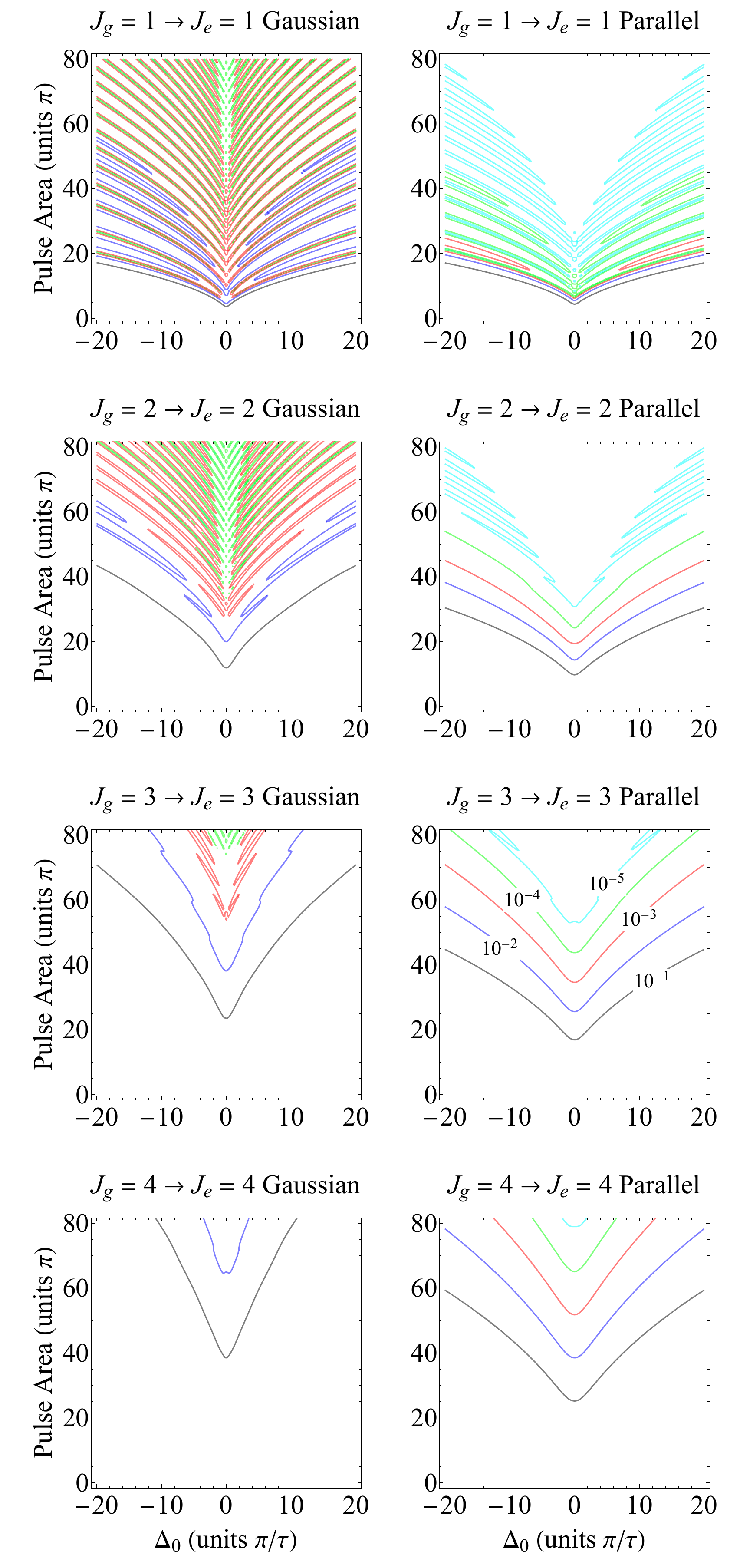}
 \caption{Two-dimensional maps of the population-transfer error $1-P_N$ as functions of the temporal pulse area and the constant single-photon detuning $\Delta_0/\pi$ for the $J_e=J_g$ family. All systems the external magnetic field is absent $B_0=0$. The left column shows the Gaussian fields of Eqs.~\eqref{eq:Gaussians}, while the right column shows the resonantly optimized parallel-STIRAP fields evaluated at the indicated detuning. The optimized pulses retain broader low-error regions, although the pulse area required for high-fidelity transfer increases with the number of states.}
 \label{fig:2d_rabi_Det}
\end{figure}


\begin{figure}[tbph]
 \centering
 \setlength{\tabcolsep}{6pt}
 \renewcommand{\arraystretch}{1.0}
 \includegraphics[width=0.9\linewidth]{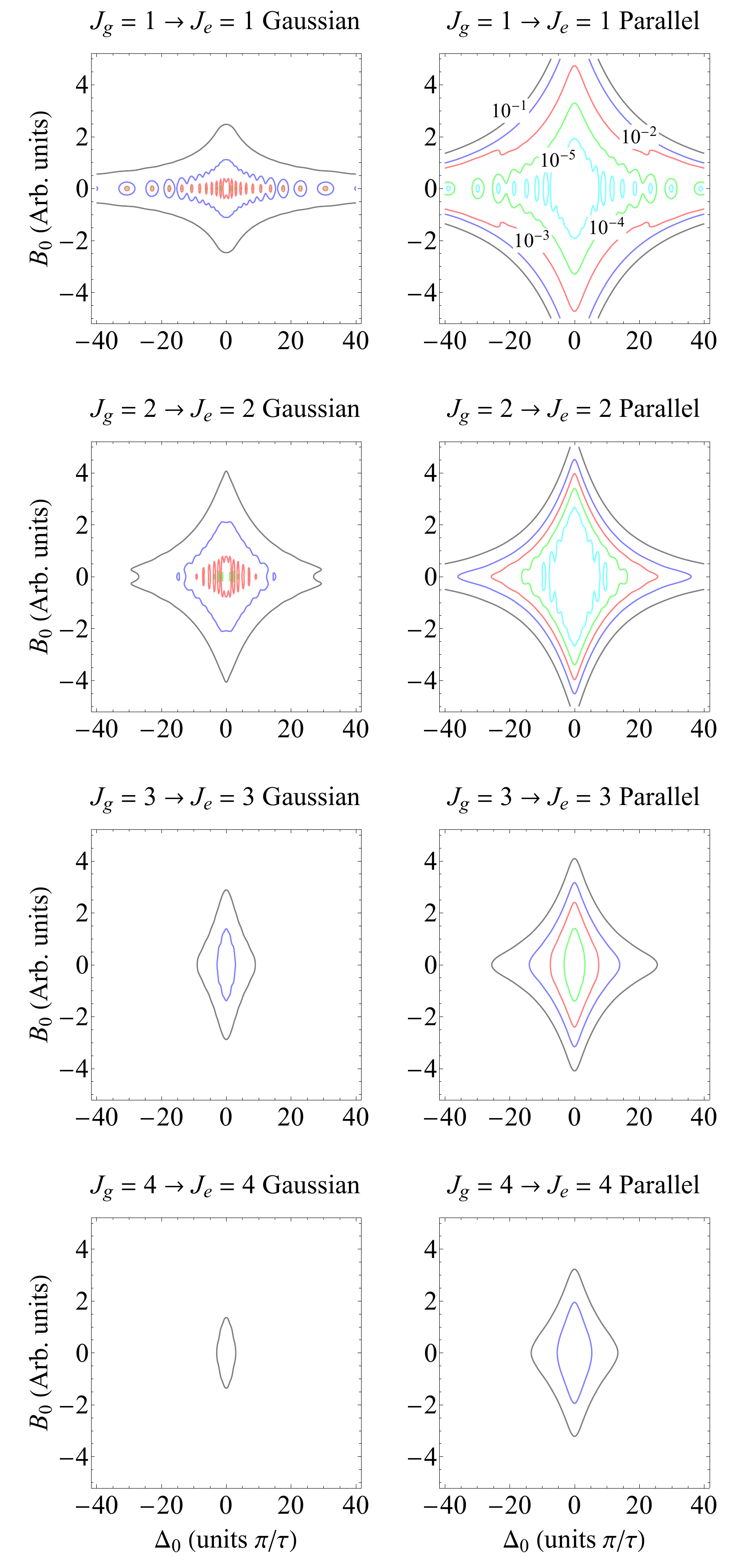}
 \label{fig:2d_det_Det}
\end{figure}

\begin{table}[tbph]
\centering
\begin{tabular}{|c|c@{\hspace{2em}}c@{\hspace{2em}}c@{\hspace{2em}}c|}
\hline
\begin{tabular}{@{}c@{}}\rule{0pt}{3ex}$J_g\to J_e$ \\[-0.5ex] g-factor\rule[-1.5ex]{0pt}{0pt}\end{tabular} &
\raisebox{0ex}{$1 \rightarrow 1$} &
\raisebox{0ex}{$2 \rightarrow 2$} &
\raisebox{0ex}{$3 \rightarrow 3$} &
\raisebox{0ex}{$4 \rightarrow 4$} \\
\hline
$g_g$ & $1$ & $1/2$ & $1/3$ & $1/4$ \\
$g_e$ & $1/3$ & $1/6$ & $1/9$ & $1/12$ \\ \hline
\end{tabular}
\caption{Land\'e g-factors for the ground $(g_g)$ and excited $(g_e)$ states which are used in the simulations for the different $J_g \rightarrow J_e$ transitions.}
\label{table:Lande g-factors}
\end{table}


For the sake of comparison, the six figures summarizing the performance of the parallel-STIRAP prescription are placed together in this section. 
Figure~\ref{fig:shapes} shows the optimized pump and Stokes pulse shapes, Fig.~\ref{fig:eigenenergies} shows the corresponding resonant adiabatic eigenenergies. Figure~\ref{fig:populations} compares the population dynamics, and Figs.~\ref{fig:1d_rabi_res}--\ref{fig:2d_det_Det} show the population transfer error as a function of the amplitude parameter $\Omega_0$, the constant single-photon detuning, and a moderate external magnetic field. 
The purpose of collecting them here is to make the comparison between the ordinary three-state problem and its multistate extensions transparent. 
Only the three-state entries are discussed in the present section; the multistate results are analyzed separately in Sec.~\ref{subsec:MultilevelSTIRAP}.

The three-state case is deliberately illustrated for both $J_g=1\rightarrow J_e=0$ and $J_g=1\rightarrow J_e=1$. 
In both cases the dynamics is that of a three-level $\Lambda$ chain, but the two choices anticipate the two angular-momentum families that appear below, namely $J\rightarrow J-1$ and $J\rightarrow J$. 
Thus the top row of the figures serves as the reference against which the multistate results will later be interpreted.

We now restrict the comparison with ordinary STIRAP to the three-state entries, i.e. to the top row of Figs.~\ref{fig:populations}--\ref{fig:2d_det_Det}. 
For Gaussian pulses \eqref{eq:Gaussians}, the total Rabi frequency $\Omega(\tau)$ changes appreciably during the same interval in which the mixing angle changes. 
Consequently, the bright-state eigenenergies are not quasiparallel in the transfer region, and leakage out of the dark state is suppressed only by increasing the pulse area.

The optimized pulses \eqref{eq:DDPpulses} with the functions \eqref{eq:1-sin2} improve this situation by concentrating the nonadiabatic coupling near the center of the interaction, where the mask $f(\tau)$ is flat and the relevant eigenenergy gap is largest.
The dark state still connects $|1\rangle$ to $|3\rangle$ as in ordinary STIRAP, but the probability of leaving this state is reduced because the eigenenergies are almost parallel in the sensitive part of the evolution.

This improvement is visible in the three-state (top) panels of Fig.~\ref{fig:1d_rabi_res}. 
At the same values of the amplitude parameter $\Omega_0$, the parallel-STIRAP pulses reach higher final-state efficiency than the Gaussian pulses, and the high-efficiency regime is reached at smaller amplitudes. 
The same conclusion is seen in the top row of Figs.~\ref{fig:2d_rabi_Det} and \ref{fig:2d_det_Det}: the optimized pulses produce broader high-fidelity regions, showing improved tolerance to variations of $\Omega_0$, to single-photon detuning, and to moderate external magnetic fields.

In Sec.~\ref{subsec:MultilevelSTIRAP} the same idea is extended to odd-numbered multistate chains, where the relevant eigenenergy gaps are additionally shaped by the Clebsch--Gordan coefficients.


\section{\label{subsec:MultilevelSTIRAP} Multistate STIRAP}

The three-state construction of Sec.~\ref{sec:theory} can be
extended naturally to chainwise-connected systems with an odd
number of states. On multiphoton resonance, such chains possess a
zero-eigenvalue dark state with no component from the excited
intermediate states, which can continuously connect the two ends
of the chain \cite{Shore1992Comparative,Vitanov1998}. Even-state
chains can also support STIRAP-like transfer, but only under
additional constraints and are not considered here.

Chainwise adiabatic passage has been studied in connection with
atomic mirrors, beam splitters, multilevel population inversion,
and population transfer between Zeeman sublevels
\cite{Marte1991,Shore1991,Smith1992,Pillet1993,Valentin1994}.
A general discussion of resonant and off-resonant chain STIRAP,
including its dressed-state interpretation, can be found in
Ref.~\cite{Vitanov-Rangelov}. In the magnetic-sublevel systems
considered below, the Clebsch--Gordan coefficients determine the
relative coupling strengths and therefore both the structure of
the dark state and its separation from the neighbouring bright
states.

\subsection{\label{subsec:multistate_model}Angular-momentum chain and Hamiltonian}


\begin{figure}
 \centering
 \includegraphics[width=0.75\linewidth]{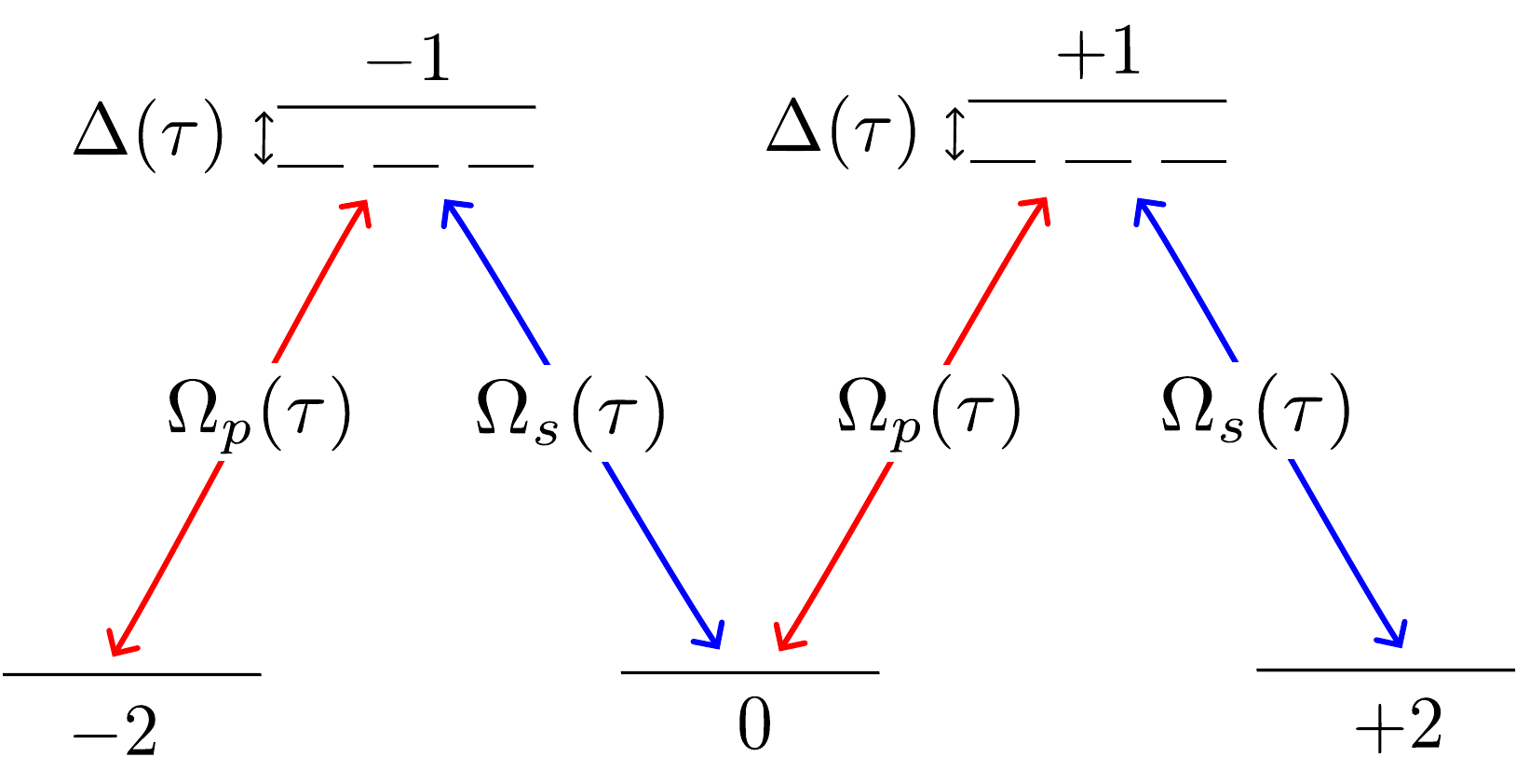}
 \caption{Five-state chainwise-connected STIRAP between magnetic sublevels of two degenerate manifolds. The chain is on four-photon resonance, while the two excited states are equally detuned by the single-photon detuning $\Delta(\tau)$.}
 \label{fig:5lvlSTIRAPgraph}
\end{figure}


We consider magnetic sublevels of two degenerate manifolds with angular momenta $J_g$ and $J_e$. 
The system is initially prepared in the state with $m_g=-J_g$, and the goal is to transfer the population to the opposite end state with $m_g=J_g$. 
Two fields with opposite circular polarizations generate an alternating chain of couplings: the pump field connects a lower-manifold sublevel to a neighbouring excited sublevel, while the Stokes field connects that excited sublevel to the next lower-manifold state. 
The resulting linkage is therefore a sequence of coupled $\Lambda$ systems, as illustrated for five states in Fig.~\ref{fig:5lvlSTIRAPgraph}.

The temporal pump and Stokes envelopes are common to all transitions, whereas the strength of each individual transition is weighted by its Clebsch--Gordan coefficient. 
Consequently, changing the angular momenta or polarization configuration modifies the adiabatic spectrum even when the applied pulse envelopes remain unchanged \cite{Shore1995,Martin1995,Martin1996}.

\begin{table}[tbph]
\centering
\begin{tabular}{|c|cccccccc|}
\hline
\begin{tabular}{@{}c@{}}\rule{0pt}{3ex}$J_g\to J_e$ \\[-0.5ex] $(\xi_{m_e}^{m_g})$\rule[-1.5ex]{0pt}{0pt}\end{tabular} &
\raisebox{0ex}{$\xi_{-3}^{-4}$} &
\raisebox{0ex}{$\xi_{-3}^{-2}$} &
\raisebox{0ex}{$\xi_{-1}^{-2}$} &
\raisebox{0ex}{$\xi_{-1}^{0}$} &
\raisebox{0ex}{$\xi_{1}^{0}$} &
\raisebox{0ex}{$\xi_{1}^{2}$} &
\raisebox{0ex}{$\xi_{3}^{2}$} &
\raisebox{0ex}{$\xi_{3}^{4}$} \\ \hline
$1\to 0$ & 0 & 0 & 0 & $\sqrt{\frac13}$ & $\sqrt{\frac13}$ & 0 & 0 & 0 \\
$1\to 1$ & 0 & 0 & 0 & $-\sqrt{\frac12}$ & $\sqrt{\frac12}$ & 0 & 0 & 0 \\ \hline
$3\to 2$ & 0 & $\sqrt{\frac57}$ & $\sqrt{\frac1{21}}$ & $\sqrt{\frac27}$ & $\sqrt{\frac27}$ & $\sqrt{\frac1{21}}$ & $\sqrt{\frac57}$ & 0 \\
$3\to 3$ & 0 & $-\frac12$ & $\sqrt{\frac5{12}}$ & $-\sqrt{\frac12}$ & $\sqrt{\frac12}$ & $-\sqrt{\frac5{12}}$ & $\frac12$ & 0 \\ \hline \hline
\begin{tabular}{@{}c@{}}\rule{0pt}{3ex}$J_g\to J_e$ \\[-0.5ex] $(\xi_{m_g}^{m_e})$\rule[-1.5ex]{0pt}{0pt}\end{tabular} &
\raisebox{0ex}{$\xi_{-4}^{-3}$} &
\raisebox{0ex}{$\xi_{-2}^{-3}$} &
\raisebox{0ex}{$\xi_{-2}^{-1}$} &
\raisebox{0ex}{$\xi_{0}^{-1}$} &
\raisebox{0ex}{$\xi_{0}^{1}$} &
\raisebox{0ex}{$\xi_{2}^{1}$} &
\raisebox{0ex}{$\xi_{2}^{3}$} &
\raisebox{0ex}{$\xi_{4}^{3}$} \\
\hline
$2\to 1$ & 0 & 0 & $\sqrt{\frac35}$ & $\sqrt{\frac1{10}}$ & $\sqrt{\frac1{10}}$ & $\sqrt{\frac35}$ & 0 & 0 \\
$2\to 2$ & 0 & 0 & $-\sqrt{\frac13}$ & $\sqrt{\frac12}$ & $-\sqrt{\frac12}$ & $\sqrt{\frac13}$ & 0 & 0 \\ \hline
$4\to 3$ & $\sqrt{\frac79}$ & $\frac16$ & $\sqrt{\frac5{12}}$ & $\sqrt{\frac16}$ & $\sqrt{\frac16}$ & $\sqrt{\frac5{12}}$ & $\frac16$ & $\sqrt{\frac79}$ \\
$4\to 4$ & $-\sqrt{\frac15}$ & $\sqrt{\frac7{20}}$ & $-\sqrt{\frac9{20}}$ & $\sqrt{\frac12}$ & $-\sqrt{\frac12}$ & $\sqrt{\frac9{20}}$ & $-\sqrt{\frac7{20}}$ & $\sqrt{\frac15}$ \\
\hline
\end{tabular}
\caption{Clebsch-Gordan coefficients for the $J_g \to J_e$ transitions considered here. The coefficients are denoted by $\xi_{m_e}^{m_g}$ for odd $J_g$ transitions (top) and $\xi_{m_g}^{m_e}$ for even $J_g$ transitions (bottom).}
\label{table:CG}
\end{table}

For the angular-momentum systems used below, the relevant chain contains $N=2J_g+1$ states. 
A convenient ordering is
$ |g_0\rangle, |e_0\rangle, |g_1\rangle, |e_1\rangle,\ldots, |e_{J_g-1}\rangle, |g_{J_g}\rangle$,
where
 $|g_r\rangle = |J_g,m_g=-J_g+2r\rangle$ ($r=0,1,\ldots,J_g$)
and $|e_r\rangle = |J_e,m_e=-J_g+2r+1\rangle$ ($r=0,1,\ldots,J_g-1$).
In this ordered basis, and in the rotating-wave approximation, the Hamiltonian reads

\bse
\begin{align}
\mathbf{H}(\tau) &= \frac{1}{2} \sum_{r=0}^{J_g-1} \Big[\Omega_{p,r}(\tau) (\vert e_r\rangle\langle g_r\vert  + \vert g_r\rangle\langle e_r\vert) \Big] \\ 
&+ \frac{1}{2} \sum_{r=0}^{J_g-1} \Big[\Omega_{s,r}(\tau) (\vert e_r\rangle\langle g_{r+1}\vert + \vert g_{r+1}\rangle\langle e_r\vert) \Big] \\
& + \sum_{r=0}^{J_g-1} \Delta(\tau) \vert e_r\rangle\langle e_r\vert  \\
& + \sum_{r=0}^{J_g-1} g_e \mu_B B_0 (2r - J_g + 1) \vert e_r\rangle\langle e_r\vert  \\
& + \sum_{r=0}^{J_g}  g_g \mu_B B_0 (2r - J_g) \vert g_r\rangle\langle g_r\vert. 
\end{align}
\label{eq:ham_Nlvl}
\ese

The first and second lines describe the pump and Stokes couplings, respectively.
The third line describes the field-induced detuning, which  in this rotating-wave picture is attributed to the upper states: Because the pump and Stokes fields are supposed to be derived from the same source, their frequency shifts are correlated and all upper states are shifted in the same manner.
The last two lines describe the Zeeman frequency shifts caused by the external magnetic field $B_0$; they are proportional to the corresponding Land\'e factors $g_g$ and $g_e$ and the corresponding magnetic quantum number, which in this notation are $m_g=2r-J_g$ and $m_e=2r-J_g+1$. 
The Bohr magnetron $\mu_B$ is set to unity throughout this study or, in other words, the magnetic field $B_0$ is measured in units $1/(\mu_B T)$.
We neglect the Stark shift from this description as it is usually much smaller than the Zeeman shift. 
In such a general situation, the single-photon and two-photon detuning lose their importance as such and the detuning dependence is more reasonably described in terms of the laser detuning $\Delta$ and the Zeeman detunings proportional to $B_0$, which is what is used in the simulations.

We note that for $B_0=0$ the multistate chain can be viewed as a sequence of two-photon resonances, as every ground state will have zero energy and all upper states will be detuned by the same detuning $\Delta$ in this picture.
This assumption will be made in the analytic derivations below, while in the figures nonzero $B_0$ will be reinstated.

The transition-dependent Rabi frequencies are given by the common pump and Stokes envelopes multiplied by the appropriate Clebsch-Gordan coefficients,
\begin{subequations}
\label{eq:multistate_couplings}
\begin{align}
 \Omega_{p,r}(\tau) &= \xi_{-J_g+2r}^{-J_g+2r+1}\,\Omega_p(\tau),
 \quad r=0,1,\ldots,J_g-1, \\
 \Omega_{s,r}(\tau) &= \xi_{-J_g+2r+2}^{-J_g+2r+1}\,\Omega_s(\tau),
 \quad r=0,1,\ldots,J_g-1 .
\end{align}
\end{subequations}
The values of these coefficients for the systems studied in this paper are listed in Table~\ref{table:CG}. 
They are the only difference between the $J_e=J_g$ and $J_e=J_g-1$ chains.

\subsection{\label{subsec:multistate_darkstate}Multistate dark state}

On multiphoton resonance, the chainwise quantum system has an instantaneous dark state with components only from the lower-manifold states,
\begin{equation}
 |D(\tau)\rangle = \frac{1}{\mathcal{N}(\tau)}
 \sum_{r=0}^{J_g} d_r(\tau)|g_r\rangle .
\label{eq:dark_multistate}
\end{equation}
The normalization factor $\mathcal{N}(\tau)$ is fixed by $\langle D(\tau)|D(\tau)\rangle=1$.
The absence of excited-state components follows from the condition $\mathbf{H}(\tau)|D(\tau)\rangle=0$. For each excited state $|e_r\rangle$ this gives the recursion relation
\begin{equation}
 \Omega_{p,r}(\tau)d_r(\tau)
 +\Omega_{s,r}(\tau)d_{r+1}(\tau)=0,
 \quad r=0,1,\ldots,J_g-1 .
\label{eq:dark_recursion}
\end{equation}
Therefore
\begin{equation}
 d_r(\tau)=(-1)^r d_0(\tau)
 \prod_{q=0}^{r-1}
 \frac{\Omega_{p,q}(\tau)}{\Omega_{s,q}(\tau)} .
\label{eq:dark_weights_general}
\end{equation}
Using Eqs.~\eqref{eq:multistate_couplings}, this can also be written as
\begin{equation}
 d_r(\tau)=(-1)^r d_0(\tau)
 \left[\frac{\Omega_p(\tau)}{\Omega_s(\tau)}\right]^r
 \prod_{q=0}^{r-1}
 \frac{\xi_{-J_g+2q}^{-J_g+2q+1}}
 {\xi_{-J_g+2q+2}^{-J_g+2q+1}} .
\label{eq:dark_weights_CG}
\end{equation}

The physical role of the counterintuitive pulse order is now transparent. At the beginning of the sequence, $\Omega_p/\Omega_s\rightarrow 0$, and the normalized dark state coincides with $|g_0\rangle$, i.e. with the initial state. 
At the end, $\Omega_s/\Omega_p\rightarrow 0$, and the same dark eigenstate coincides with $|g_{J_g}\rangle$, i.e. with the target end state. If the evolution follows this dark state adiabatically, population transfer is achieved without populating the intermediate excited sublevels.

The existence and structure of this state are standard results for odd chainwise-connected systems \cite{Marte1991,Shore1991,Smith1992,Vitanov1998}. For example, in a five-state chain the unnormalized dark state has the form
\begin{equation}
 |D(\tau)\rangle \propto
 \Omega_{2,3}\Omega_{4,5}|1\rangle
 -\Omega_{1,2}\Omega_{4,5}|3\rangle
 +\Omega_{1,2}\Omega_{3,4}|5\rangle ,
\end{equation}
which explicitly shows the absence of the upper states $|2\rangle$ and $|4\rangle$.

\subsection{\label{sec:OptMultiSTIRAP}Pulse shapes for five-state STIRAP}

The pulse shapes for five states are derived from the condition to make the closest-to-zero eigenenergies parallel.
It is shown in Appendix \ref{appendix} that these eigenenergies are given by the expressions
\bse
\begin{align}
\ev_{\text{nearest},\pm}^{(2\rightarrow 1)}(\tau) &= \pm\frac{\Omega(\tau)}{2} \sqrt{\frac{7-\sqrt{1+24\zeta(\tau)^2}}{20}}, \\
\ev_{\text{nearest},\pm}^{(2\rightarrow 2)}(\tau) &= \pm\frac{\Omega(\tau)}{2} \sqrt{\frac{5-\sqrt{9-8\zeta(\tau)^2}}{12}},
\end{align}
\ese
where 
\be
\zeta(\tau) = \cos 2\theta.
\ee
These expressions show that the quasiparallel mask function $f(\tau)$ is not sufficient to make the above eigenenergies quasiparallel due to the time dependence of the square root caused by $\zeta(\tau)$. 
Such a time dependence was absent in the three-state case; hence the function $f(\tau)$ was sufficient.

In order to make these eigenenergies quasiparallel, we use the same mask function $f(\tau)$ as for three states, but introduce an additional function $\lambda(\tau)$,
\bse
\label{eq:multiPulses}
\begin{align}
 \Omega_p(\tau)&=\fac\Omega_0 \lambda(\tau) f(\tau)\sin\theta(\tau),\\
 \Omega_s(\tau)&=\fac\Omega_0 \lambda(\tau) f(\tau)\cos\theta(\tau),
\end{align}
\label{eq:shapes_multi}
\ese
which is the normalized inverse of the square root, so that $\lambda^{(J_g\rightarrow J_e)}(0)=1$.
These formulas apply for seven and nine states below, although with different expressions for $\fac$ and $\lambda(\tau)$.

In the two linkage cases we have
\bse
\begin{align}
\lambda^{(2\rightarrow 1)}(\tau) &= \sqrt{\frac{6}{7-\sqrt{1+24\zeta(\tau)^2}}}, \\
\lambda^{(2\rightarrow 2)}(\tau) &= \sqrt{\frac{2}{5-\sqrt{9-8\zeta(\tau)^2}}}.
\end{align}
\ese
Substitution into the nearest eigenenergies gives
\bse
\begin{align}
\ev_{\text{nearest},\pm}^{(2\to1)}(\tau) &=
\pm \frac{\kappa_{2\to1}\Omega_0}{2} \sqrt{\frac{3}{10}}\, f(\tau),
\\
\ev_{\text{nearest},\pm}^{(2\to2)}(\tau) &= \pm \frac{\kappa_{2\to2}\Omega_0}{2\sqrt{6}}\,f(\tau).
\end{align}
\ese
Thus, the dependence on the mixing angle is cancelled exactly, while the mask $f(\tau)$ forces the eigenenergies to vanish smoothly at the beginning and end of the interaction.

The factors $\kappa_{J_g\to J_e}$ are chosen so that each individual pump and Stokes pulse has area $\pi\Omega_0$.
They are therefore determined by
\be
\kappa_{J_g\to J_e}
=
\frac{\pi}{
\displaystyle
\int_{-\infty}^{+\infty}
\lambda^{(J_g\to J_e)}(\tau)
f(\tau)
\sin\theta(\tau)d\tau
}.
\ee
The corresponding Stokes-pulse integral has the same value by symmetry. 
Numerical integration gives $\kappa_{2\to1}\approx1.584656$ and $\kappa_{2\to2}\approx2.900627$.
For $\Omega_0=1$, the corresponding central dark-bright gaps are
$\left|\eta^{(2\rightarrow1)}_{\mathrm{nearest},\pm}(0)\right| \approx0.433976$ and
$\left|\eta^{(2\rightarrow2)}_{\mathrm{nearest},\pm}(0)\right| \approx0.592088$.

\subsection{Pulse shapes for seven-state STIRAP}

For the seven-state chains, the characteristic equation for
the squared nonzero eigenenergies is cubic.
The explicit cubic polynomials and their analytic roots are
given in Appendix \ref{appendix}. The bright-state eigenenergies closest
to the dark state are
\bse
\begin{align}
\ev_{\mathrm{nearest},\pm}^{(3\to2)}(\tau)
&= \pm\frac{\Omega(\tau)}{2} \left\{\frac{22}{63} + \frac{2}{63} \sqrt{13+138\zeta^2(\tau)} \right. \notag\\
&\times \left.\cos\left[\phi_{3\to2}(\tau)+\frac{2\pi}{3}\right]\right\}^{1/2},
\\
\ev_{\mathrm{nearest},\pm}^{(3\to3)}(\tau)
&= \pm\frac{\Omega(\tau)}{2} \left\{\frac{7}{18}+\frac{1}{18} \sqrt{7\left[7-6\zeta^2(\tau)\right]} \right. \notag\\
&\times \left.\cos\left[\phi_{3\to3}(\tau)+\frac{2\pi}{3}\right]\right\}^{1/2},
\end{align}
\ese
where
\bse
\begin{align}
\phi_{3\to3}(\tau)
&=
\frac{1}{3}
\arccos\left[\frac{143-153\zeta^2(\tau)}{\left\{7\left[7-6\zeta^2(\tau)\right]\right\}^{3/2}}\right].
\\
\phi_{3\to2}(\tau)
&= \frac{1}{3}
\arccos\left[
\frac{909\zeta^2(\tau)-35}{\left[13+138\zeta^2(\tau)\right]^{3/2}}\right].
\end{align}
\ese

The centre-normalized parallelizing functions are therefore
defined by
\bse
\begin{align}
\lambda^{(3\to2)}(\tau)
&= \sqrt{\frac{5}{21}} \left\{\frac{22}{63} + \frac{2}{63} \sqrt{13+138\zeta^2(\tau)} \right. \notag\\
&\times \left.\cos\left[\phi_{3\to2}(\tau)+\frac{2\pi}{3}\right]\right\}^{-1/2},
\\
\lambda^{(3\to3)}(\tau)
&= \frac{1}{\sqrt{12}} \left\{\frac{7}{18}+\frac{1}{18} \sqrt{7\left[7-6\zeta^2(\tau)\right]} \right. \notag\\
&\times \left.\cos\left[\phi_{3\to3}(\tau)+\frac{2\pi}{3}\right]\right\}^{-1/2},
\end{align}
\ese
which satisfy $\lambda^{(3\to2)}(0)=\lambda^{(3\to3)}(0)=1$.

The pump and Stokes fields retain the forms of Eqs.~\eqref{eq:multiPulses}.
Consequently, the nearest eigenenergies reduce to
\bse
\begin{align}
\ev_{\text{nearest},\pm}^{(3\to2)}(\tau)
&= \pm \frac{\kappa_{3\to2}\Omega_0}{2}
\sqrt{\frac{5}{21}}\,f(\tau),
\\
\ev_{\text{nearest},\pm}^{(3\to3)}(\tau)
&= \pm \frac{\kappa_{3\to3}\Omega_0}{4\sqrt{3}}\,f(\tau).
\end{align}
\ese
Hence, as in the five-state case, the parallelizing function
removes the mixing-angle dependence of the closest
dark--bright gap exactly.

Requiring each pump and Stokes pulse to have area
$\pi\Omega_0$ gives
$\kappa_{3\to2}\approx1.372502$,
and
$\kappa_{3\to3} \approx 3.367581$.
For $\Omega_0=1$, the corresponding central dark-bright
gaps are
$\left|\ev_{\mathrm{nearest},\pm}^{(3\to2)}(0)\right|
\approx 0.334856$ and
$\left|\ev_{\mathrm{nearest},\pm}^{(3\to3)}(0)\right|
\approx0.486068$.

\begin{table}[t]
\centering
\caption{Splitting between the null dark energy and the closest bright eigenenergies for the optimized pulses at
$\tau=0$ and $\Omega_0=1$. The values are given in units of $1/\tau$.}
\label{tab:eigenvalue_splitting}
\begin{tabular}{|c|c|c|}
\hline
$N$ & Transition & Dark-bright splitting \\
\hline
3 & $J_g=1\rightarrow J_e=0$ & 0.673929 \\
3 & $J_g=1\rightarrow J_e=1$ & 0.825392 \\
5 & $J_g=2\rightarrow J_e=1$ & 0.433976 \\
5 & $J_g=2\rightarrow J_e=2$ & 0.592088 \\
7 & $J_g=3\rightarrow J_e=2$ & 0.334856 \\
7 & $J_g=3\rightarrow J_e=3$ & 0.486068 \\
9 & $J_g=4\rightarrow J_e=3$ & 0.278156 \\
9 & $J_g=4\rightarrow J_e=4$ & 0.421866 \\
\hline
\end{tabular}
\end{table}

\begin{figure}
    \centering
    \includegraphics[width=0.6\linewidth]{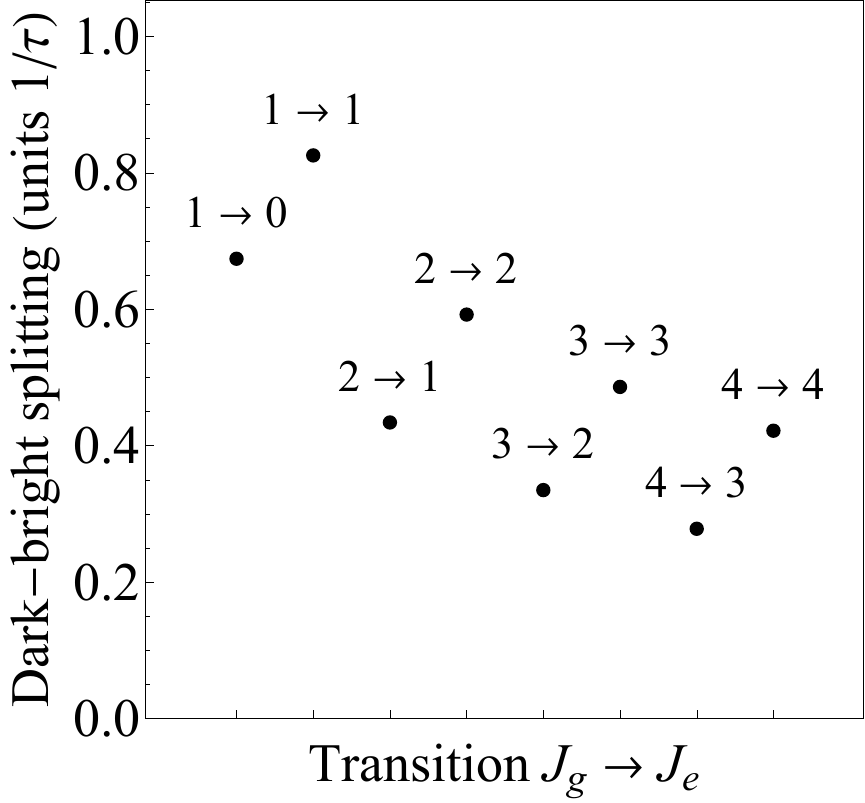}
    \caption{Dark-bright state splitting for the optimized pulses. The data points correspond to the values tabulated in Table \ref{tab:eigenvalue_splitting}, representing the splitting between the null dark energy and the nearest bright eigenenergies for various transitions $J_g \to J_e$.}
    \label{fig:Energy_splittings}
\end{figure}

\subsection{Pulse shapes for nine-state STIRAP}

For the nine-state chains, the characteristic equation for
the squared nonzero eigenenergies is quartic. We label the two
linkages by $\ell\in\{4\to3,4\to4\}$ and denote the smallest
positive solution of the corresponding quartic by
$\mathcal{R}_{\ell}(\tau)$. Its explicit derivation is given
in Appendix \ref{appendix}. In terms of the quartic coefficients, it can
be written as
\begin{align}
\mathcal{R}_{\ell}(\tau)
&= -\frac{a_{\ell}}{4} -\frac{1}{2}
\Bigg[\sqrt{m_{\ell}(\tau)} \notag\\
&+\sqrt{-2\alpha_{\ell}(\tau)-m_{\ell}(\tau) + \frac{2\beta_{\ell}(\tau)}{\sqrt{m_{\ell}(\tau)}}}\,
\Bigg].
\end{align}
Here $m_{\ell}(\tau)$ is the largest positive root of the
resolvent cubic. Introducing
\bse
\begin{align}
\mathcal{P}_{\ell}(\tau)
 &= -\frac{\alpha_{\ell}^{2}(\tau)}{3} -4\gamma_{\ell}(\tau),
\\
\mathcal{Q}_{\ell}(\tau)
&= -\frac{2\alpha_{\ell}^{3}(\tau)}{27} +\frac{8\alpha_{\ell}(\tau)\gamma_{\ell}(\tau)}{3} -\beta_{\ell}^{2}(\tau),
\end{align}
\ese
this root is
\begin{align}
m_{\ell}(\tau) &= -\frac{2\alpha_{\ell}(\tau)}{3} + 2\sqrt{-\frac{\mathcal{P}_{\ell}(\tau)}{3}} \notag \\
&\times \cos\left[ \frac{1}{3} \arccos\left( \frac{3\mathcal{Q}_{\ell}(\tau)}{2\mathcal{P}_{\ell}(\tau)}
\sqrt{-\frac{3}{\mathcal{P}_{\ell}(\tau)}}\,\right)
\right].
\end{align}
The functions $a_{\ell}$, $\alpha_{\ell}(\tau)$,
$\beta_{\ell}(\tau)$, and $\gamma_{\ell}(\tau)$ are listed
explicitly in Appendix \ref{appendix}.

The bright-state eigenenergies closest to the dark state are
therefore
\bse
\begin{align}
\ev_{\mathrm{nearest},\pm}^{(4\to3)}(\tau)
&= \pm\frac{\Omega(\tau)}{2}\sqrt{\mathcal{R}_{4\to3}(\tau)},
\\
\ev_{\mathrm{nearest},\pm}^{(4\to4)}(\tau)
&= \pm\frac{\Omega(\tau)}{2}\sqrt{\mathcal{R}_{4\to4}(\tau)}.
\end{align}
\ese
The normalized parallelizing functions are therefore
\bse
\begin{align}
\lambda^{(4\to3)}(\tau) &=\sqrt{\frac{7}{36\mathcal{R}_{4\to3}(\tau)}},
\\
\lambda^{(4\to4)}(\tau) &=\frac{1}{\sqrt{20\mathcal{R}_{4\to4}(\tau)}},
\end{align}
\ese
which satisfy
$\lambda^{(4\to3)}(0)=\lambda^{(4\to4)}(0)=1$.

The pump and Stokes fields retain the forms of Eqs.~\eqref{eq:multiPulses}.
Substitution of the parallelizing functions into the nearest eigenenergies gives
\bse
\begin{align}
\ev_{\mathrm{nearest},\pm}^{(4\to3)}(\tau) &=\pm \frac{\sqrt{7}}{12} \kappa_{4\to3}\Omega_0 f(\tau),
\\
\ev_{\mathrm{nearest},\pm}^{(4\to4)}(\tau) &=\pm\frac{1}{4\sqrt{5}}\kappa_{4\to4}\Omega_0 f(\tau).
\end{align}
\ese
Thus, as in the five- and seven-state cases, the parallelizing function cancels the mixing-angle dependence of the closest dark--bright gap exactly.

Requiring each pump and Stokes pulse to have area
$\pi\Omega_0$ gives $\kappa_{4\to3}\approx1.261595$ and
$\kappa_{4\to4}\approx3.773281$. For $\Omega_0=1$, the
corresponding central dark--bright gaps are
$\left|\ev_{\mathrm{nearest},\pm}^{(4\to3)}(0)\right|\approx0.278156$,
and $\left|\ev_{\mathrm{nearest},\pm}^{(4\to4)}(0)\right| \approx 0.421866$.

The eigenenergy gaps for all angular-momentum systems with three to nine states are listed in Table~\ref{tab:eigenvalue_splitting} and illustrated in Fig.~\ref{fig:Energy_splittings}.
The values are evaluated at $\tau=0$ and $\Omega_0=1$, and represent the distance between the zero-energy dark state and either of its two nearest bright neighbours.

\subsection{\label{subsec:multistate_trends}Numerical results}

The optimized pulse shapes and their corresponding resonant eigenenergy spectra are shown in Figs.~\ref{fig:shapes} and \ref{fig:eigenenergies}. 
By construction, the parallelizing function $\lambda(\tau)$ removes the mixing-angle dependence of the eigenenergies closest to the dark state. 
Their remaining time dependence is therefore determined  only by the mask $f(\tau)$, which is chosen to have a flat top in a broad temporal range. 
Consequently, the relevant dark-bright gap is nearly constant in the central interaction region, where the dark state changes most rapidly, and vanishes smoothly at the pulse boundaries.
We observe that the parallelization condition leads to somewhat unusual pulse shapes, albeit still in the counterintuitive order Stokes-pump.

Only the pair of eigenenergies nearest to the dark state is parallelized. 
This is sufficient because these states provide the dominant channels for nonadiabatic leakage. 
The remaining bright eigenenergies retain a nontrivial time dependence, as visible in Fig.~\ref{fig:eigenenergies}, but remain farther from the dark state throughout the central transfer region.

Two systematic trends are evident. 
First, the central gap decreases as the number of states increases as seen in Figure \ref{fig:Energy_splittings}. 
A longer chain  contains more eigenstates, and the eigenenergy spectrum near the dark state becomes progressively denser. 
Adiabatic following therefore requires a larger pulse area as $N$ increases. 
Second, for every value of $N$, the $J_e=J_g$ linkage has a larger dark-bright gap than the corresponding $J_e=J_g-1$ linkage. 
This difference originates from the more balanced Clebsch-Gordan coefficients (cf. Table \ref{table:CG}), which determine the effective couplings.

These spectral trends directly explain the resonant transfer results in Fig.~\ref{fig:1d_rabi_res}. 
At the same temporal pulse area, the parallel-STIRAP pulses produce more stable state dynamics as shown in Fig.~\ref{fig:populations} and a smaller transfer error than the Gaussian reference for all chain lengths. 
The improvement can reach several orders of magnitude in the adiabatic regime. 
Furthermore, the $J_e=J_g$ systems generally enter the high-efficiency regime at smaller pulse areas than the corresponding $J_e=J_g-1$ systems, consistently with their larger central dark-bright gaps.

The increase in the required pulse area with $N$ is not removed by the optimization. 
Rather, the parallelizing function makes more effective use of the available pulse area by keeping the smallest gap large and nearly constant during the part of the evolution in which the nonadiabatic coupling is strongest. 
The improvement is therefore most naturally understood as a suppression of local nonadiabatic leakage.

For clarity, the robustness maps in Figs.~\ref{fig:2d_rabi_Det} and \ref{fig:2d_det_Det} are shown only for the more favourable $J_e=J_g$ family. 
Figure~\ref{fig:2d_rabi_Det} displays the transfer error as a function of the temporal pulse area and the constant single-photon detuning. 
As expected, the pulse area required for high-fidelity transfer increases with both the detuning and the number of states. 
Nevertheless, the optimized pulses produce much broader high-efficiency regions and require a smaller pulse area than the Gaussian pulses over the range of detunings considered.

Figure~\ref{fig:2d_det_Det} shows the combined sensitivity to the single-photon detuning and the end-state $(N-1)$-photon detuning at fixed $\Omega_0$. 
The optimized pulses again greatly broaden the low-infidelity region around exact multiphoton resonance. 
The  tolerance nevertheless decreases with increasing $N$, reflecting the progressively smaller separation between the dark state and its nearest bright neighbours as $N$ increases, as seen in Table \ref{tab:eigenvalue_splitting}.

The two detunings affect the transfer differently. 
A common single-photon detuning shifts the nonzero eigenvalues up or down, thereby decreasing the splitting from the null dark energy and hence deteriorating the transfer efficiency.
However, the single-photon detuning preserves the chain of two-photon resonances and hence maintains the existence of the dark state. 
By contrast, the multiphoton detuning directly perturbs the dark-state condition and mixes the dark state with neighbouring bright states, as in three-state systems~\cite{Vitanov-Rangelov}. 
It is therefore particularly detrimental in longer chains. 
The quasiparallel construction reduces this sensitivity, but it cannot remove the requirement of approximate resonance.

Overall, the numerical results confirm that the transfer efficiency is governed primarily by the eigenenergy gap in the region where the dark state evolves most rapidly. 
Flattening this gap substantially improves both the resonant transfer and its robustness against amplitude and detuning variations, while the Clebsch-Gordan structure makes the $J_e=J_g$ family consistently more favourable.


\section{\label{sec:dissipations}Effects of dissipation in three- and five-state STIRAP}

The preceding sections considered coherent dynamics. We now test whether the quasiparallel pulse construction remains useful when the excited states are lossy. This is important because the central advantage of STIRAP is not only adiabatic following, but adiabatic following through a dark state with small excited-state population. Dissipation therefore probes a different aspect of the optimization: it checks whether suppressing nonadiabatic transitions also reduces the time-integrated population in the decaying manifold.

\subsection{Master-equation model}

We describe spontaneous emission from the excited sublevels with the Gorini-Kossakowski-Sudarshan-Lindblad master equation~\cite{Lindblad1976,Gorini-Kassakowski-Sudarsahan1976}. In the dimensionless time variable used throughout the paper, the density matrix obeys
\begin{equation}
 \dot{\rho}(\tau)
 = -i\left[H(\tau),\rho(\tau)\right]
 + \sum_n
 \left[
 C_n \rho(\tau) C_n^{\dagger}
 -\frac{1}{2}\left\{C_n^{\dagger}C_n,\rho(\tau)\right\}
 \right].
 \label{eq:gksl_dissipation}
\end{equation}
Here $H(\tau)$ is the coherent Hamiltonian: Eq.~\eqref{eq:ham_3lvl} for the three-state system and Eq.~\eqref{eq:ham_Nlvl} for the multistate chain. The sum runs over all allowed decay channels from the excited magnetic sublevels to the ground-state manifold.

For a decay channel $\ket{m_e}\rightarrow\ket{m_g}$, the collapse operator is written as
 $C_{m_e\rightarrow m_g} = \sqrt{\gamma_{m_e\rightarrow m_g}}\, A_{m_e\rightarrow m_g}$,
where
$A_{m_e\rightarrow m_g}=|m_g\rangle\langle m_e|$.
The partial rates are weighted by the dipole-transition strengths.
For a decay channel returning to a ground sublevel retained in the chain, we use
\begin{equation}
 \gamma_{m_e\rightarrow m_g} = \Gamma\left|\xi^{m_e}_{m_g}\right|^2,
 \label{eq:partial_decay_rate}
\end{equation}
where the Clebsch-Gordan coefficients are normalized so that the strengths of all dipole-allowed decay channels from a fixed $m_e$ sum to unity. 
Decays to ground sublevels outside the selected STIRAP chain, including omitted polarization channels, are represented by an uncoupled sink state $|\ell_{m_e}\rangle$ with rate
\begin{equation}
 \gamma_{m_e\rightarrow\ell_{m_e}} =
 \Gamma\Bigg[ 1-\sum_{m_g\in\mathrm{chain}} \left|\xi^{m_e}_{m_g}\right|^2 \Bigg].
 \label{eq:loss_decay_rate}
\end{equation}
The Hamiltonian acts trivially on the sink states and population entering them does not return to the driven chain. Equations~\eqref{eq:partial_decay_rate} and \eqref{eq:loss_decay_rate} therefore preserve the total decay rate $\Gamma$ for every excited sublevel while treating external branches as irreversible loss.

\subsection{Numerical comparison}

\begin{figure}[tbph]
 \begin{tabular}{c}
 \includegraphics[width=0.9\linewidth]{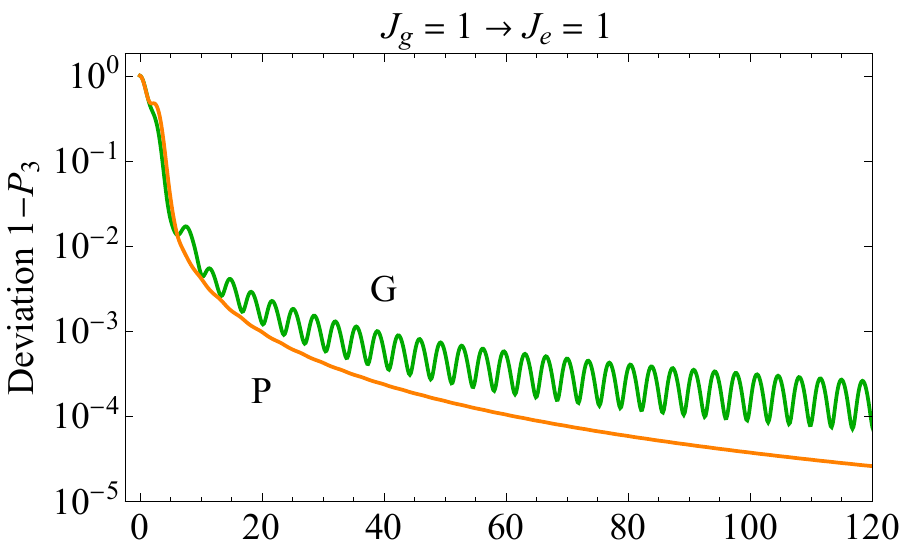} \\
 \includegraphics[width=0.9\linewidth]{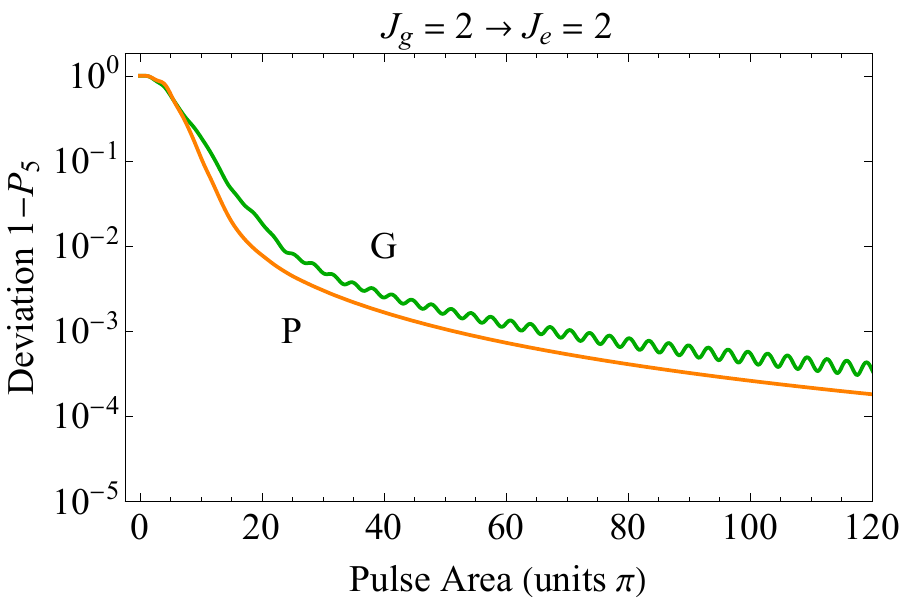}
 \end{tabular}
 \caption{Population-transfer error $1-P_N$ as a function of the temporal pulse area in the presence of spontaneous emission from the excited manifold. Gaussian pulses are denoted by G and parallel-STIRAP with P. The dimensionless decay rate is $\Gamma=0.5$. The top and bottom panels show the $J_g=1\rightarrow J_e=1$ three-state chain and the $J_g=2\rightarrow J_e=2$ five-state chain, respectively. 
 } 
 \label{fig:dissipations}
\end{figure}

We compare the two pulse choices shown in Fig.~\ref{fig:dissipations}: ordinary resonant Gaussian pulses and resonant pulses optimized for quasiparallel process. 
The comparison is made for the favorable $J_e=J_g$ chains, namely the three-state $J_g=1\rightarrow J_e=1$ system and the five-state $J_g=2\rightarrow J_e=2$ system. In all cases the decay scale is fixed to dimensionless $\Gamma=0.5$ i.e. in units $1/\tau$.

The results in Fig.~\ref{fig:dissipations} show that the optimized resonant pulses retain a clear advantage over the Gaussian reference in the presence of spontaneous emission.
The reason is the same as in the closed-system analysis, but with an additional consequence. 
Quasiparallel eigenenergies reduce leakage from the dark state to the bright states. Since the bright states contain excited-state components, this also lowers the time-integrated population exposed to spontaneous emission. 
Thus the pulse shaping improves the final transfer not by adding a new decay-avoidance mechanism, but by making the usual STIRAP dark-state following more accurate.

\section{\label{sec:conclusion} Conclusion}

We have developed a pulse-shaping method for improving adiabatic population transfer in multistate STIRAP chains with an odd number of states. 
The method extends the quasiparallel-eigenenergy approach, previously used in quantum systems with two and three states to multistate STIRAP. 
Importantly, it modifies only the shapes of the pump and Stokes pulses already required for STIRAP and introduces neither auxiliary shortcut fields nor direct couplings between distant states.

The method is based on the smallest nonzero eigenvalue of the Hamiltonian, which determines the eigenenergy separation between the dark state and its nearest bright neighbours. 
A common time-dependent amplitude factor is chosen to cancel the mixing-angle dependence of this closest eigenvalue, which is the dominant nonadiabatic leakage channel. 
Since this factor leaves the null vector of the coupling matrix unchanged, the dark-state trajectory is preserved while the relevant dark-bright gap is made nearly constant in the central part of the interaction, where nonadiabatic transitions are most likely.

We derived analytic pulse shapes for three-, five-, seven-, and nine-state chains formed by magnetic sublevels of degenerate angular-momentum manifolds. 
Numerical simulations show that the resulting parallel-STIRAP pulses can reduce the transfer error by several orders of magnitude compared with conventional Gaussian pulses of the same temporal area.
The optimized pulses also produce broader high-fidelity regions with respect to variations of the peak Rabi frequencies, the single-photon detuning, and the moderate external magnetic field.

The results reveal two systematic trends. 
First, the dark-bright gap decreases as the number of states increases, so larger pulse areas are required to reach the same transfer accuracy in longer chains. 
Pulse shaping does not remove this spectral crowding, but it makes more efficient use of the available coupling strength by flattening the smallest gap during the relevant part of the evolution. 
Second, the $J_e=J_g$ chains consistently outperform the corresponding $J_e=J_g-1$ chains. 
This difference originates from their Clebsch--Gordan coefficients, which determine the effective couplings and hence the separation between the dark state and the nearest bright eigenstate.

We also examined spontaneous emission from the excited manifold in representative three- and five-state systems. 
For the decay rate considered, the optimized resonant pulses retain a clear advantage over the Gaussian reference. 
The same suppression of nonadiabatic leakage that improves the closed-system transfer also reduces the population entering the lossy bright states.

These results demonstrate that engineering the local adiabatic spectrum provides a simple and effective route to high-fidelity multistate STIRAP. 
The method is particularly suitable for systems in which additional shortcut couplings are unavailable or difficult to implement. 
Moreover, the angular-momentum systems considered here can be driven by light delivered from the same laser source and polarized with orthogonal circular $\sigma_\pm$ polarizations.

Natural extensions include its application to other odd chain geometries, numerical optimization of the mask and mixing functions, and the incorporation of platform-specific constraints such as bounded amplitudes, finite bandwidth, unequal couplings, and additional decoherence mechanisms.


\acknowledgments
This research is supported by the Bulgarian national plan for recovery and resilience, Contract No. BG-RRP-2.004-0008-C01 (SUMMIT), Project No. 3.1.4, by the European Union’s Horizon Europe research and innovation program under Grant Agreement No. 101046968 (BRISQ), and the QuantERA project FALCON. 
We acknowledge the use of the supercomputing cluster PhysOn at Sofia University for this work.


\appendix

\section{Direct derivation of the instantaneous eigenenergies\label{appendix}}
\label{sec:direct_eigenvalues}

The instantaneous eigenenergies can be obtained directly from the characteristic polynomial of Eq.~\eqref{eq:ham_Nlvl},
$\det \left[\ev \mathbf{I} - (\mathbf{H}) \right] = 0$. 
On resonance, $\Delta(\tau)=0$, and in the ordered basis $\{|g_0\rangle,|e_0\rangle,|g_1\rangle,\ldots, |e_{J_g-1}\rangle,|g_{J_g}\rangle\}$, the Hamiltonian is tridiagonal, with consecutive couplings $p_r/2$ and $s_r/2$, where $p_r=a_r\Omega_p$, $s_r=b_r\Omega_s$, $a_r=\xi^{-J_g+2r+1}_{-J_g+2r}$, and $b_r=\xi^{-J_g+2r+1}_{-J_g+2r+2}$. 
The coefficients $a_r$ and $b_r$ are listed in Table~\ref{table:CG}.

Let $\mathcal D_n(\ev)$ denote the characteristic determinant of the upper-left $n\times n$ block. 
For consecutive off-diagonal matrix elements $h_1=p_0/2$, $h_2=s_0/2$, $h_3=p_1/2,\ldots$, the determinant obeys
\begin{align}
&\mathcal D_0=1,\quad
\mathcal D_1=\ev,\quad \ldots \notag \\
&\mathcal D_n=\ev\mathcal D_{n-1}-h_{n-1}^2\mathcal D_{n-2}.
\label{eq:direct_recursion}
\end{align}
Only the squared couplings enter Eq.~\eqref{eq:direct_recursion}; therefore, the signs of the Clebsch--Gordan coefficients do not affect the eigenenergies. 

Because the chain contains an odd number of states and has zero diagonal elements, its characteristic polynomial contains one zero root $\ev_0=0$ and $J_g$ pairs of opposite roots. 
For compactness, define
$\zeta=(\Omega_s^2-\Omega_p^2)/\Omega^2=\cos 2\theta$ and $x=4\ev^2/\Omega^2$. 
After inserting the Clebsch-Gordan coefficients from Table~\ref{table:CG}, the characteristic equation reduces to
\begin{equation}
\mathcal D_{2J_g+1}(\ev)
\propto
\ev P_{J_g\rightarrow J_e}(x,\zeta).
\label{eq:reduced_direct_characteristic}
\end{equation}
If $x_j(\zeta)$ are the positive roots of $P_{J_g\rightarrow J_e}(x,\zeta)$, the instantaneous spectrum is
\begin{equation}
\ev_0(\tau)=0,\quad
\ev_{k,\pm}(\tau)
=
\pm\frac{\Omega(\tau)}{2}\sqrt{x_k[\zeta(\tau)]}.
\label{eq:direct_spectrum}
\end{equation}
The bright eigenvalue closest to the dark state is therefore obtained from the smallest positive root $x_{\min}(\zeta)$.

\subsection{Three states}

For the three-state chains, direct evaluation gives $P_{1\rightarrow0}=3x-1$ and $P_{1\rightarrow1}=2x-1$. 
The eigenvalues are
\be
\ev_0(\tau)=0,\quad 
\ev_{j,\pm}^{(1\rightarrow 0)}(\tau) = \pm\frac{\Omega(\tau)}{2}\sqrt{\frac13},
\ee
for $J_g=1\rightarrow J_e=0$, and 
\be
\ev_0(\tau)=0,\quad 
\ev_{j,\pm}^{(1\rightarrow 1)}(\tau) = \pm\frac{\Omega(\tau)}{2}\sqrt{\frac12},
\ee
for $J_g=1\rightarrow J_e=1$.

\subsection{Five states}

For the five-state chain, the recursion gives
\begin{equation}
\mathcal D_5
=
\ev\left[
\ev^4
-\frac{p_0^2+s_0^2+p_1^2+s_1^2}{4}\ev^2
+\frac{p_0^2p_1^2+p_0^2s_1^2+s_0^2s_1^2}{16}
\right].
\end{equation}
For the $J_g=2\rightarrow J_e=1$ linkage,
after substitution of the corresponding coefficients, we find
$P_{2\rightarrow 1}(x,\zeta) = 50x^2-35x+6-3\zeta^2$,
with roots
\be
x_{1,2}^{(2\rightarrow 1)}(\zeta) = \frac{7\mp\sqrt{1+24\zeta^2}}{20},
\ee

For the $J_g=2\rightarrow J_e=2$ linkage, we have
$P_{2\rightarrow 2}(x,\zeta) = 18x^2-15x+2+\zeta^2$,
with roots
\be
x_{1,2}^{(2\rightarrow 2)}(\zeta) = \frac{5\mp\sqrt{9-8\zeta^2}}{12},
\ee
The corresponding five instantaneous eigenenergies are
\bse
\begin{gather}
\ev_0(\tau) = 0,
\\
\ev_{k,\pm}(\tau)
=
\pm\frac{\Omega(\tau)}{2}
\sqrt{x_k[\zeta(\tau)]},
\quad k=1,2.
\end{gather}
\ese
The eigenenergies closest to $\ev_0(\tau)$ are
\bse
\begin{align}
\ev_{1,\pm}^{(2\rightarrow 1)}(\tau) &= \pm\frac{\Omega(\tau)}{2} \sqrt{\frac{7-\sqrt{1+24\zeta(\tau)^2}}{20}}, \\
\ev_{1,\pm}^{(2\rightarrow 2)}(\tau) &= \pm\frac{\Omega(\tau)}{2} \sqrt{\frac{5-\sqrt{9-8\zeta(\tau)^2}}{12}}.
\end{align}
\ese

\subsection{Seven states}

Repeated use of the tridiagonal recurrence gives, for the
$J_g=3\to J_e=2$ linkage,
\be
P_{3\to2}(x,\zeta) = 3087x^3 - 3234x^2 + \left(1099-322\zeta^2\right)x + 90\zeta^2 - 120.
\ee
For the $J_g=3\to J_e=3$ linkage, one obtains
\be
P_{3\to3}(x,\zeta) = 288x^3 - 336x^2 + \left(98+28\zeta^2\right)x - 6 - 9\zeta^2.
\ee

For the $J_g=3\to J_e=2$ linkage, define
\be
\phi_{3\to2}(\zeta) = \frac{1}{3}\arccos\left[\frac{909\zeta^2-35}{\left(13+138\zeta^2\right)^{3/2}}\right].
\ee
The three positive roots are ($k=0,1,2$)
\be
x_k^{(3\to2)}(\zeta) = \frac{22}{63} + \frac{2}{63} \sqrt{13+138\zeta^2}
\cos\left[ \phi_{3\to2}(\zeta) - \frac{2\pi k}{3} \right].
\ee

For the $J_g=3\to J_e=3$ linkage, define
\be
\phi_{3\to3}(\zeta)=\frac{1}{3}\arccos\left[\frac{143-153\zeta^2}{\left[7\left(7-6\zeta^2\right)\right]^{3/2}}\right].
\ee
The three positive roots are  ($k=0,1,2$)
\be
x_k^{(3\to3)}(\zeta)=\frac{7}{18}+\frac{1}{18}\sqrt{7\left(7-6\zeta^2\right)}
\cos\left[\phi_{3\to3}(\zeta)-\frac{2\pi k}{3}\right].
\ee

For the physical interval $-1\leq\zeta\leq1$, the smallest
root in both cases corresponds to $k=2$. Hence,
\be
x_{\min}^{(3\to2)}(\zeta) = \frac{22}{63} + \frac{2}{63} \sqrt{13+138\zeta^2}
\cos\left[\phi_{3\to2}(\zeta) + \frac{2\pi}{3}\right]
\ee
and
\be
x_{\min}^{(3\to3)}(\zeta) = \frac{7}{18} + \frac{1}{18}\sqrt{7\left(7-6\zeta^2\right)}
\cos\left[\phi_{3\to3}(\zeta) + \frac{2\pi}{3}\right].
\ee

The seven instantaneous eigenenergies are
\bse
\begin{gather}
\ev_0(\tau)=0,
\\
\ev_{k,\pm}^{(J_g\to J_e)}(\tau)
= \pm\frac{\Omega(\tau)}{2} \sqrt{x_k^{(J_g\to J_e)}[\zeta(\tau)]},
\quad k=0,1,2.
\end{gather}
\ese

\subsection{Nine states}

Repeated use of Eq.~\eqref{eq:direct_recursion} gives for the
$J_g=4\to J_e=3$ and  $J_g=4\to J_e=4$ linkages,
\bse
\begin{align}
P_{4\to3}(x,\zeta)
&= 23328x^4-32400x^3 +(16434-3348\zeta^2)x^2
\notag\\
&+ \left(1983\zeta^2-3582\right)x +35\zeta^4-280\zeta^2+280,
\\
P_{4\to4}(x,\zeta)
&= 20000x^4-30000x^3 +\left(13650+2700\zeta^2\right)x^2
\notag\\
&-\left(2050+1755\zeta^2\right)x +27\zeta^4+216\zeta^2+72.
\end{align}
\ese

For either linkage, division by the coefficient of $x^4$
gives the monic quartic ($\ell\in\{4\to3,4\to4\}$)
\be
x^4+a_{\ell}x^3+b_{\ell}(\zeta)x^2
+c_{\ell}(\zeta)x+d_{\ell}(\zeta)=0.
\ee

For the $J_g=4\to J_e=3$ linkage, the coefficients are
\bse
\begin{align}
a_{4\to3} & =-\frac{25}{18}, \\
b_{4\to3}(\zeta) &= \frac{913-186\zeta^2}{1296}, \\
c_{4\to3}(\zeta) &=\frac{-1194+661\zeta^2}{7776}, \\
d_{4\to3}(\zeta) &= \frac{35(\zeta^4-8\zeta^2+8)}{23328}.
\end{align}
\ese
For the $J_g=4\to J_e=4$ linkage, they are
\bse
\begin{align}
a_{4\to4} & =-\frac{3}{2}, \\
b_{4\to4}(\zeta) &= \frac{273+54\zeta^2}{400}, \\
c_{4\to4}(\zeta) &= -\frac{410 + 351\zeta^2}{4000}, \\
d_{4\to4}(\zeta) &= \frac{9(3\zeta^4+24\zeta^2+8)}{20000}.
\end{align}
\ese

To solve the quartic, we first remove the cubic term by
introducing the standard depressed-quartic coefficients
\bse
\begin{gather}
\alpha_{\ell}
=
b_{\ell}-\frac{3a_{\ell}^{2}}{8},
\quad
\beta_{\ell}
=
c_{\ell}-\frac{a_{\ell}b_{\ell}}{2}
+\frac{a_{\ell}^{3}}{8},
\\
\gamma_{\ell}
=
d_{\ell}
-\frac{a_{\ell}c_{\ell}}{4}
+\frac{a_{\ell}^{2}b_{\ell}}{16}
-\frac{3a_{\ell}^{4}}{256}.
\end{gather}
\ese
Their arguments $\zeta$ are omitted here for compactness.

For the $J_g=4\to J_e=3$ linkage, these functions become
\bse
\begin{gather}
\alpha_{4\to3}(\zeta)=-\frac{49+372\zeta^2}{2592},
\quad
\beta_{4\to3}(\zeta)=\frac{1-19\zeta^2}{1296},
\\
\gamma_{4\to3}(\zeta)=\frac{5\left(77+1128\zeta^2+8064\zeta^4\right)}{26873856}.
\end{gather}
\ese

For the $J_g=4\to J_e=4$ linkage, one finds
\bse
\begin{gather}
\alpha_{4\to4}(\zeta)=\frac{3\left(36\zeta^2-43\right)}{800},
\quad
\beta_{4\to4}(\zeta)=\frac{27\zeta^2-25}{2000},
\\
\gamma_{4\to4}(\zeta)=\frac{3\left(1547-2664\zeta^2+1152\zeta^4\right)}{2560000}.
\end{gather}
\ese

The corresponding resolvent equation is
\be
m^3+2\alpha_{\ell}m^2 + \left(\alpha_{\ell}^{2}-4\gamma_{\ell}\right)m -\beta_{\ell}^{2} = 0.
\ee
Introducing
\be
\mathcal{P}_{\ell}
=
-\frac{\alpha_{\ell}^{2}}{3}
-4\gamma_{\ell},
\quad
\mathcal{Q}_{\ell}
=
-\frac{2\alpha_{\ell}^{3}}{27}
+\frac{8\alpha_{\ell}\gamma_{\ell}}{3}
-\beta_{\ell}^{2},
\ee
the largest positive root of the resolvent cubic is
\be
m_{\ell}
=
-\frac{2\alpha_{\ell}}{3}
+
2\sqrt{-\frac{\mathcal{P}_{\ell}}{3}}
\cos\left\{
\frac{1}{3}
\arccos\left[
\frac{3\mathcal{Q}_{\ell}}
{2\mathcal{P}_{\ell}}
\sqrt{-\frac{3}{\mathcal{P}_{\ell}}}
\right]
\right\}.
\ee

For convenience, define $S_{\ell}=\sqrt{m_{\ell}}$ and
\be
T_{\ell,\pm}
=
\sqrt{
-2\alpha_{\ell}-m_{\ell}
\pm\frac{2\beta_{\ell}}{S_{\ell}}
}.
\ee
The four roots of the quartic are then
\bse
\begin{gather}
x_1^{(\ell)}
=
-\frac{a_{\ell}}{4}
-\frac{S_{\ell}+T_{\ell,+}}{2},
\quad
x_2^{(\ell)}
=
-\frac{a_{\ell}}{4}
-\frac{S_{\ell}-T_{\ell,+}}{2},
\\
x_3^{(\ell)}
=
-\frac{a_{\ell}}{4}
+\frac{S_{\ell}-T_{\ell,-}}{2},
\quad
x_4^{(\ell)}
=
-\frac{a_{\ell}}{4}
+\frac{S_{\ell}+T_{\ell,-}}{2}.
\end{gather}
\ese

Throughout the physical interval $-1\leq\zeta\leq1$, the
smallest positive root is $x_1^{(\ell)}(\zeta)$. The nine
instantaneous eigenenergies are
\bse
\begin{gather}
\ev_0(\tau)=0,
\\
\ev_{k,\pm}^{(\ell)}(\tau)
=
\pm\frac{\Omega(\tau)}{2}
\sqrt{x_k^{(\ell)}[\zeta(\tau)]},
\quad
k=1,2,3,4.
\end{gather}
\ese
Consequently, the eigenenergies closest to the dark state are
\be
\ev_{\mathrm{nearest},\pm}^{(\ell)}(\tau)
=
\pm\frac{\Omega(\tau)}{2}
\sqrt{x_1^{(\ell)}[\zeta(\tau)]}.
\ee

At the pulse centre, $\zeta(0)=0$. For the
$J_g=4\to J_e=3$ linkage, the polynomial factorizes as
\be
P_{4\to3}(x,0)
=
2(3x-1)(9x-4)(12x-5)(36x-7).
\ee
Its four roots are $7/36$, $1/3$, $5/12$, and $4/9$;
therefore,
$x_1^{(4\to3)}(0)=\frac{7}{36}$.
For the $J_g=4\to J_e=4$ linkage, one obtains
\be
P_{4\to4}(x,0) = 2(5x-4)(5x-1)(20x-9)(20x-1).
\ee
The four roots are $1/20$, $1/5$, $9/20$, and $4/5$, so
that $x_1^{(4\to4)}(0)=\frac{1}{20}$.

\section{Pulse noise effects on infidelity for three- and five-state STIRAP}
\label{app:noisy_pulses}

\begin{figure}[tbph]
 \includegraphics[width=0.9\linewidth]{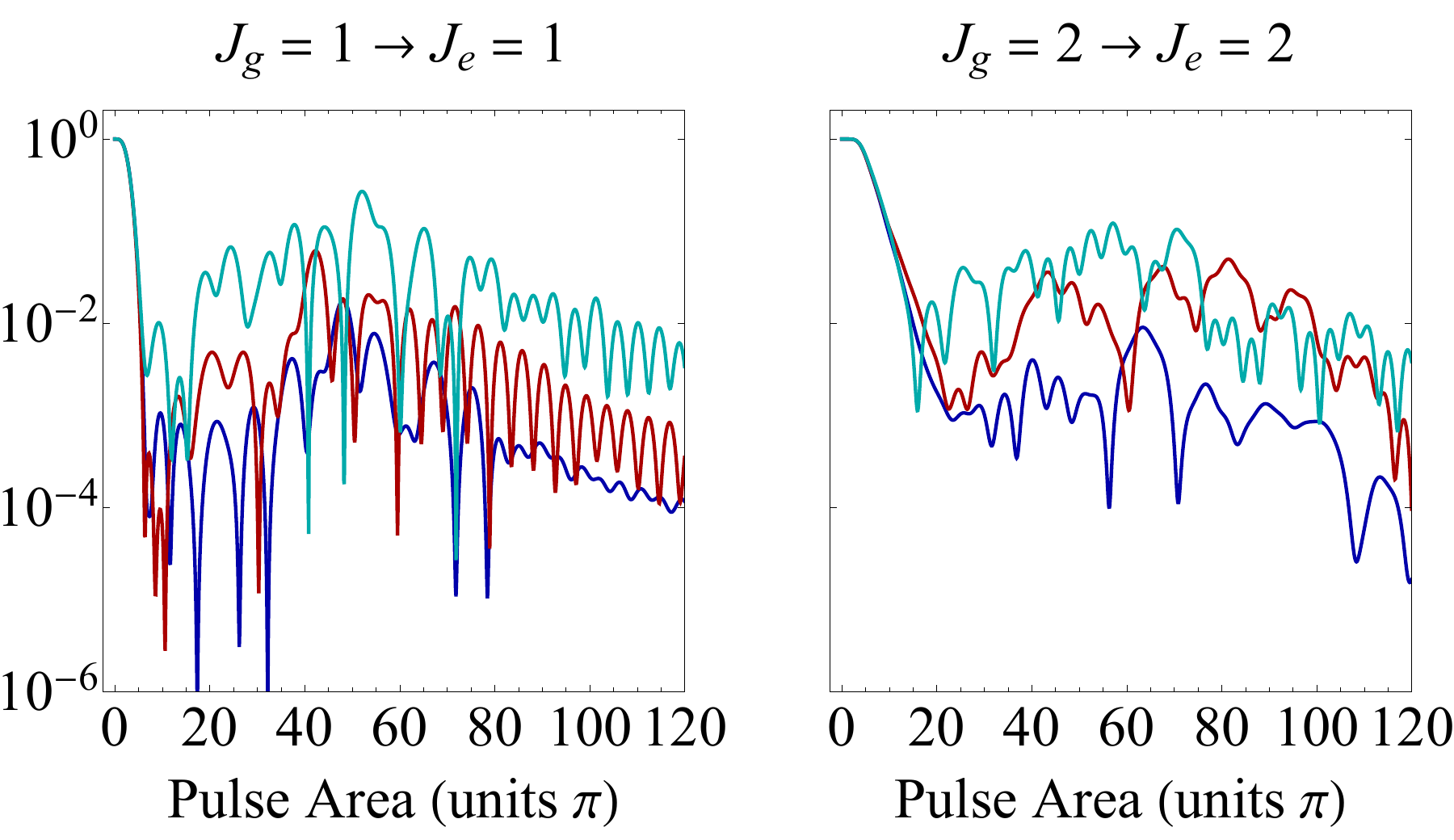}
 \caption{Population-transfer error $1-P_N$ as a function of the temporal pulse area in the presence of white noise in the optimal pulses with amplitude of $1\%$ (dark blue), $2\%$ (dark red), $5\%$ (cyan). Simulations are run for three-state $J_g=1 \rightarrow J_e=1$ and five-state $J_g=2 \rightarrow J_e=2$ STIRAP both on resonance and in the absence of external magnetic fields.
 } 
 \label{fig:noisy_fidelity}
\end{figure}

Figure \ref{fig:noisy_fidelity} illustrates the final-state infidelity for both the three- and five-state resonant STIRAP under the influence of noisy driving fields. To quantify the robustness against experimental imperfections, we introduce amplitude fluctuations to the laser pulses at relative magnitudes of $1\%$ (dark blue), $2\%$ (dark red), and $5\%$ (cyan). As anticipated, deviations from the optimal analytic pulse profiles in Fig. \ref{fig:shapes} disrupt the adiabatic evolution, leading to a corresponding decline in the overall population transfer efficiency. At sufficiently large pulse areas, the inherent robustness of the adiabatic passage begins to suppress the dynamic errors induced by these shape fluctuations. Under these strongly coupled conditions, the system reaches steady infidelity decline, rendering the transfer dynamics largely insensitive to minor local field noise. Specifically, for the three-state ($J_g=1 \rightarrow J_e=1$) linkage, this noise-resilient regime is achieved at pulse areas of approximately $80\pi$.

\section{Dephasing}
\label{app:dephasing}

We extend the adiabatic treatment of dephasing in three-state STIRAP by Ivanov \textit{et al.} \cite{PeshoDephasing} to the odd chains considered here. The essential difference is that dephasing-induced transfer out of the dark state generally feeds several bright eigenstates. The dark-state population therefore does not obey a closed scalar equation, and the natural reduced description is a rate equation for the populations of the instantaneous adiabatic states.

\subsection{Adiabatic-state kinetics}

Let $\{|j\rangle\}$ denote the bare states of a chain and assume pure dephasing that is diagonal in this basis. A completely positive Markovian model is:
\bse
\begin{align}
 \dot\rho&=-\frac{i}{\hbar}[H,\rho]
 +\sum_\mu\left(L_\mu\rho L_\mu^\dagger
 -\frac12\{L_\mu^\dagger L_\mu,\rho\}\right),
 \label{eq:deph-master}\\
 L_\mu&=\sum_j \ell_{\mu j}|j\rangle\langle j|.
\end{align}
\ese
The bare coherence $\rho_{jk}$ decays at the rate
\begin{equation}
 \gamma_{jk}=\frac12\sum_\mu|\ell_{\mu j}-\ell_{\mu k}|^2.
 \label{eq:bare-deph-rate}
\end{equation}
Thus only differences between the noise shifts of two states cause dephasing; a common shift has no effect.

Let $U(t)$ have the instantaneous eigenstates $|a(t)\rangle$ of $H(t)$ as its columns, and define $\rho_a=U^\dagger\rho U$. The exact equation in this basis is
\bse
\begin{align}
 \dot\rho_a&=-i[H_a,\rho_a]
 +\sum_\mu\mathcal D[L_{\mu,a}]\rho_a,\\
 H_a&=U^\dagger H U-i U^\dagger\dot U,
 \qquad L_{\mu,a}=U^\dagger L_\mu U,
 \label{eq:adiabatic-master}
\end{align}
\ese
where $\mathcal D[L]\rho=L\rho L^\dagger-\{L^\dagger L,\rho\}/2$. The second term in $H_a$ is the geometric coupling between adiabatic states. When the evolution is adiabatic and dephasing is weak compared with the nonzero adiabatic gaps, coherences between nondegenerate adiabatic states can be eliminated to leading order. The populations $q_a=\langle a|\rho|a\rangle$ then satisfy
\bse
\begin{align}
 \dot q_a&=\sum_{b\ne a}
 \left(\Gamma_{b\rightarrow a}q_b-\Gamma_{a\rightarrow b}q_a\right),
 \label{eq:adiabatic-rates}\\
 \Gamma_{b\rightarrow a}&=\sum_\mu
 |\langle a|L_\mu|b\rangle|^2.
 \label{eq:adiabatic-rate}
\end{align}
\ese
For degenerate or nearly degenerate adiabatic levels, the coherences within the corresponding subspace must instead be retained in Eq.~(\ref{eq:adiabatic-master}).

For an initially pure dark state $|D\rangle$, the instantaneous rate of transfer out of it is
\begin{align}
 R_D&=\sum_{a\ne D}\Gamma_{{D\rightarrow a}}
 =\sum_\mu\left[\langle D|L_\mu^\dagger L_\mu|D\rangle
 -|\langle D|L_\mu|D\rangle|^2\right].
 \label{eq:dark-variance}
\end{align}
For real diagonal $L_\mu$, this is the sum of their variances in $|D\rangle$. Equivalently,
\begin{equation}
 R_D=2\sum_{j<k}\gamma_{jk}p_jp_k,
 \qquad p_j=|\langle j|D\rangle|^2.
 \label{eq:dark-pair-rate}
\end{equation}
In the weak-dephasing limit, the leading transfer loss is $1-P_{\rm tar}\simeq\int R_D(t)\,dt$. Return from bright states and higher-order dephasing corrections are included by solving Eq.~(\ref{eq:adiabatic-rates}).

\subsection{Five-state $J_g=2\to J_e=2$ chain}

For the chain $|g_{-2}\rangle,|e_{-1}\rangle,|g_0\rangle,|e_{+1}\rangle,|g_{+2}\rangle$, the magnitudes of the four Clebsch--Gordan factors are $1/\sqrt3$, $1/\sqrt2$, $1/\sqrt2$, and $1/\sqrt3$. In a phase convention with positive couplings, the Hamiltonian is
\begin{equation}
 H=\frac{1}{2}
 \begin{pmatrix}
 0&\Omega_p/\sqrt3&0&0&0\\
 \Omega_p/\sqrt3&2\Delta&\Omega_s/\sqrt2&0&0\\
 0&\Omega_s/\sqrt2&0&\Omega_p/\sqrt2&0\\
 0&0&\Omega_p/\sqrt2&2\Delta&\Omega_s/\sqrt3\\
 0&0&0&\Omega_s/\sqrt3&0
 \end{pmatrix}.
 \label{eq:J22-hamiltonian}
\end{equation}
On multiphoton resonance it has the dark state
\begin{equation}
 |D\rangle=\mathcal{N}^{-1}(\tau)\left(|g_{-2}\rangle-\sqrt{2/3}x|g_0\rangle
 +x^2|g_{+2}\rangle\right),
 \label{eq:J22-dark}
\end{equation}
where $x=\Omega_p(\tau)/\Omega_s(\tau)$ and $\mathcal{N}(\tau)=\left(1+2x^2/3+x^4\right)^{1/2}$.
Its lower-manifold populations are
\begin{equation}
 p_{-2}=\mathcal{N}^{-2}(\tau),\qquad p_0=\frac{2x^2}{3\mathcal{N}^2(\tau)},\qquad
 p_{+2}=\frac{x^4}{\mathcal{N}^2(\tau)};
 \label{eq:J22-pops}
\end{equation}
the excited-state populations vanish.

Consider magnetic-field noise described by a diagonal operator $Z_m$ with eigenvalues $z_j$ in the bare basis,
\begin{equation}
 \dot\rho\big|_{\rm B}=-\gamma_B[Z_m,[Z_m,\rho]].
 \label{eq:magnetic-dephasing}
\end{equation}
This convention gives $\gamma_{jk}=\gamma_B(z_j-z_k)^2$ and the adiabatic transition rates
\begin{equation}
 \Gamma_{b\rightarrow a}=2\gamma_B
 |\langle a|Z_m|b\rangle|^2.
 \label{eq:J22-ad-rates}
\end{equation}
If the three ground sublevels have $z_{g_m}=g_gB_0m$, the initial dark-state leakage rate is
\begin{align}
 R_D&=2\gamma_B g_g^2 B_0^2
 \left[4(p_{-2}+p_{+2})-4(p_{+2}-p_{-2})^2\right] \notag \\
 &=\frac{16}{3}\gamma_Bg_g^2 B_0^2
 \frac{x^2+6x^4+x^6}{(1+2x^2/3+x^4)^2}.
 \label{eq:J22-leak}
\end{align}
The equivalent pairwise form is
\begin{equation}
 R_D=2\gamma_Bg_g^2 B_0^2\left[
 4p_{-2}p_0+4p_0p_{+2}+16p_{-2}p_{+2}\right].
 \label{eq:J22-pairwise}
\end{equation}
The $m=0$ state has no magnetic phase shift relative to zero, but it does contribute through its coherences with $m=\pm2$. Its role is explicit in Eq.~(\ref{eq:J22-pairwise}); its dependence is hidden in Eq.~(\ref{eq:J22-leak}) only because $p_{-2}+p_0+p_{+2}=1$.

At equal pump and Stokes amplitudes, $x=1$, the dark-state populations are $(p_{-2},p_0,p_{+2})=(3/8,1/4,3/8)$ and $R_D=6\gamma_Bg_g^2B_0^2$. The two pairs involving $m=0$ contribute $3\gamma_Bg_g^2 B_0^2/2$ in total, while the $(-2,+2)$ pair contributes $9\gamma_Bg_g^2 B_0^2/2$. The leakage vanishes at either pulse tail, where the dark state becomes a single bare state.

The magnetic coefficient in Eq.~(\ref{eq:magnetic-dephasing}) can accommodate different Land\'e factors in the two manifolds by using the actual $z_j$ for each state. Equation~(\ref{eq:J22-leak}) uses a common ground-manifold factor $g_g^2 B_0^2$ only because the dark state has support on ground states alone. Excited-state noise still affects subsequent redistribution among bright states through Eq.~(\ref{eq:J22-ad-rates}).

\subsection{Implications for the parallelized pulses}

The pulse shapes in Eqs.~\eqref{eq:shapes_multi} multiply $\sin\theta$ and $\cos\theta$ by the same factor $\kappa\Omega_0\lambda(\tau)f(\tau)$. On exact single-photon resonance and at fixed $\theta(\tau)$ the Hamiltonian is multiplied by a common scalar, so its instantaneous eigenvectors are unchanged. In this case all rates in Eq.~(\ref{eq:adiabatic-rate}) are unchanged: parallelization suppresses nonadiabatic leakage by increasing and flattening the relevant gaps, but does not by itself suppress Markovian pure-dephasing transitions. At nonzero single-photon detuning, the detuning is not scaled with the fields, and the bright eigenvectors and their dephasing-induced redistribution rates can change. The direct dark-state leakage rate in Eq.~(\ref{eq:dark-pair-rate}) nevertheless remains unchanged at fixed $\theta(\tau)$, since the dark state has no excited-state component and depends only on the pump-to-Stokes ratios. Dephasing performance can also change if pulse shaping changes this dark-state trajectory or the physical duration. In dimensionless time $\tau=t/T$, the dephasing strength enters as $\gamma_BT$.

The rate description assumes adiabatic following and weak dephasing. Where the dark--bright gap becomes small, or when coherences in degenerate subspaces matter, the full master equation (\ref{eq:adiabatic-master}) should be integrated. These qualifications distinguish dephasing-induced transfer loss from the spontaneous-emission effects treated separately in the main text.

\begin{figure}[H]
\centering
 \includegraphics[width=0.9\linewidth]{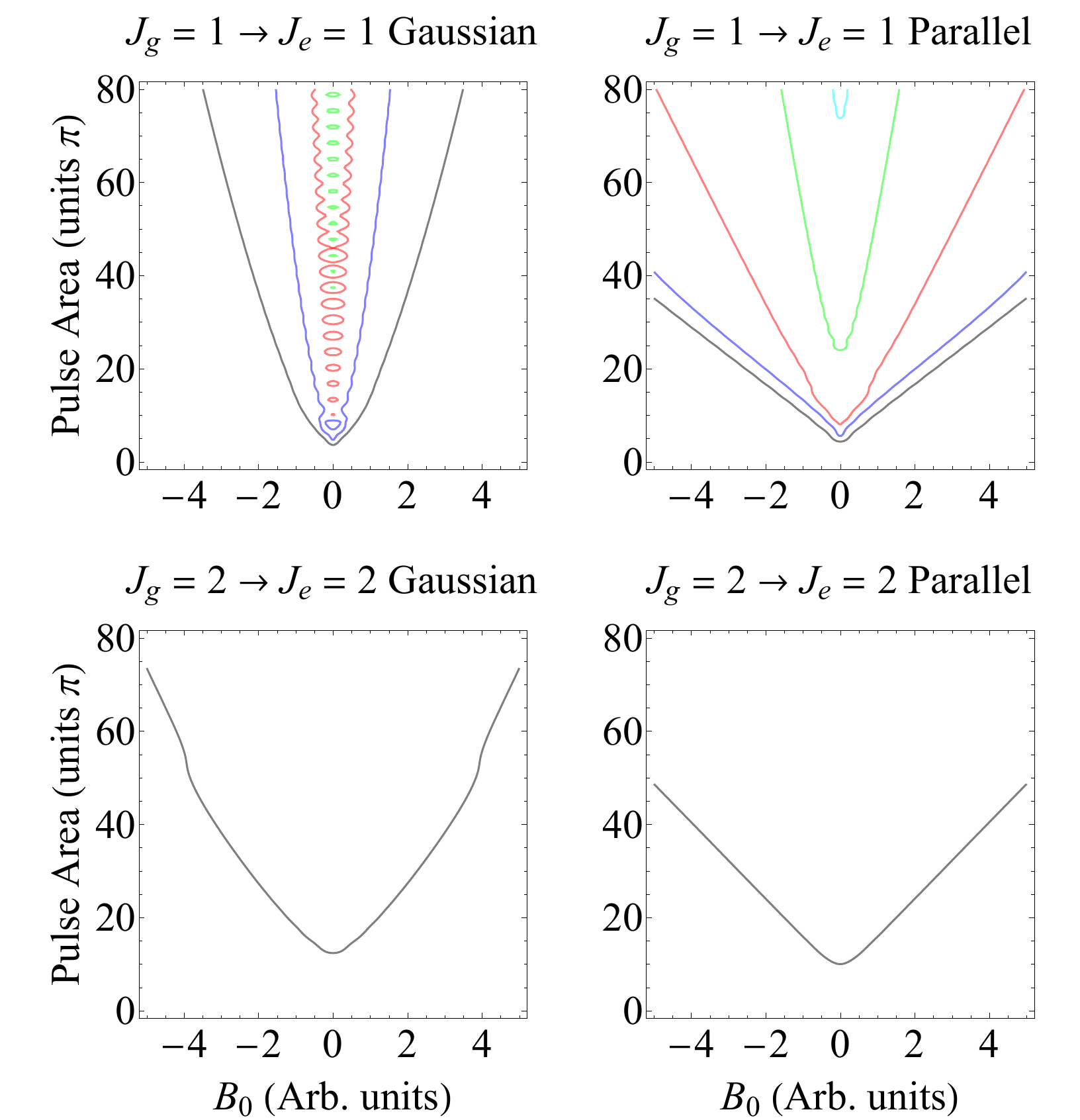}
 \caption{Population-transfer error $1-P_N$ as a function of the temporal pulse area and the external magnetic field strength $B_0$ in the presence of ground state dephasing with rate $\gamma_B = 0.05$. Simulations are run for three-state $J_g=1 \rightarrow J_e=1$ and five-state $J_g=2 \rightarrow J_e=2$ resonant STIRAP. The respective Land\'e g-factors are listed in Table~\ref{table:Lande g-factors}.} 
 \label{fig:dephasing}
\end{figure}

Fidelity is greatly affected by dephasing and deteriorates significantly with the number states as can be seen in Fig.~\ref{fig:dephasing}. Optimized fields \eqref{eq:shapes_multi} still outperform the Gaussian pulses by making three- and five-state STIRAP with dephasing robust against the external magnetic field $B_0$ while still requiring lower pulse area in order to reach the same fidelity.

\end{document}